\documentclass[compsoc,conference,a4paper,10pt,times]{IEEEtran}
\IEEEoverridecommandlockouts

\usepackage{cite}
\usepackage{amsmath,amssymb,amsfonts}
\usepackage{graphicx}
\usepackage{textcomp}
\usepackage{xcolor}
\usepackage{epsfig,endnotes}
\usepackage{url}
\usepackage{breakurl}
\usepackage[breaklinks]{hyperref}
\usepackage{multirow}
\usepackage{balance}
\usepackage{hyperref}
\usepackage{algorithm,algpseudocode}
\usepackage{subfigure}
\usepackage{tablefootnote}
\usepackage{booktabs}
\usepackage{multirow}
\usepackage{placeins}
\usepackage{color}
\usepackage{tablefootnote}
\usepackage{placeins}
\usepackage{tabu}
\usepackage{tikz}
\usetikzlibrary{plotmarks}
\usepackage{flafter}
\usepackage{moresize}

\newcommand{\agan}{{\emph{Jekyll}}} %
\newcommand{\eg}{{\em e.g.,\ }}
\newcommand{\ie}{{\em i.e.\ }}

\newcommand{\todo}[1]{{\color{red} #1}}

\newcommand{\NA}[1]{{\todo{N/A}}}
\newcommand{\para}[1]{{\vspace{1pt} \bf \noindent #1 \hspace{6pt}}}

\makeatletter
\newcommand\notsotiny{\@setfontsize\notsotiny{6.9}{7.1828}}
\makeatother

\def\BibTeX{{\rm B\kern-.05em{\sc i\kern-.025em b}\kern-.08em
    T\kern-.1667em\lower.7ex\hbox{E}\kern-.125emX}}
\begin{document}

\def\UrlBreaks{\do\/\do-}

\title{Jekyll: Attacking Medical Image Diagnostics using Deep Generative Models}

\author{
 {\rm Neal Mangaokar\textsuperscript{1}, Jiameng Pu\textsuperscript{1}, Parantapa Bhattacharya\textsuperscript{2}, 
 Chandan K. Reddy\textsuperscript{1}, Bimal Viswanath\textsuperscript{1} }\\ 
  \textsuperscript{1}Virginia Tech \hspace{0.05in}
 \textsuperscript{2}University of Virginia \hspace{0.05in} \\
 {\small \{neal1998, jmpu, ckreddy, vbimal\}@vt.edu, parantapa@virginia.edu}
 }

\maketitle

\begin{abstract}
Advances in deep neural networks (DNNs) have shown tremendous promise in the medical domain. However, the deep learning tools that are helping the domain, can also be used against it. Given the prevalence of fraud in the healthcare domain, it is important to consider the adversarial use of DNNs in manipulating sensitive data that is crucial to patient healthcare. In this work, we present the design and implementation of a DNN-based image translation attack on biomedical imagery. More specifically, we propose \agan{}, a neural style transfer framework that takes as input a biomedical image of a patient and translates it to a new image that indicates an attacker-chosen disease condition. The potential for fraudulent claims based on such generated `fake' medical images is significant, and we demonstrate successful attacks on both X-rays and retinal fundus image modalities. We show that these attacks manage to mislead both medical professionals and algorithmic detection schemes. Lastly, we also investigate defensive measures based on machine learning to detect images generated by \agan{}.

\end{abstract}

\section{Introduction}
\label{sec:intro}

\noindent As we make rapid advances in deep learning/AI, it is important to understand the associated security implications. Attackers can no longer be assumed to have limited algorithmic intelligence~\cite{brundage2018malicious}. An emerging threat is that of ``deepfakes'', or AI generated synthetic content that appears convincingly real~\cite{cnndeepfakes}. Deepfakes are enabled by deep generative models such as Generative Adversarial Neural Networks (GANs)~\cite{goodfellow2014generative}. Generative models can produce photorealistic fake images~\cite{karras2017progressive, karras2018style}, and convincing fake videos~\cite{vondrick2016generating}. Recently, the threat of deepfakes has been largely discussed in the context of the Web, where they can be used to create fake accounts, fake pornographic images of celebrities, images of people doing things they never did to spread misinformation, and manipulate elections~\cite{vicebjpdeepfake}. In this work, we investigate threats posed by deepfakes in the healthcare domain, \ie how bad actors can make use of generative schemes to attack critical workflows in our healthcare framework.

Healthcare spending is huge in many developed countries, and the system is already fraught with fraud~\cite{finlayson2019adversarial, al2018fraud, morris2009combating, rashidian2012no, bauder2017medicare, herland2018big}. Prior work has highlighted exploitable vulnerabilities in the healthcare domain. The industry has employed poor security practices in securing sensitive patient data such as biomedical images~\cite{beek_2018, verizon_2018, mirsky2019ctgan}. In 2019, Mirsky et al. demonstrated attacks that compromise biomedical data management pipelines~\cite{mirsky2019ctgan}. As our healthcare system is susceptible to bad actors, it is important to investigate new threat vectors driven by technological advances.

We propose an attack framework called \agan{} that leverages generative deep learning to derail one of the most important decision processes in the medical domain---\textit{medical diagnosis based on biomedical image analysis}. \agan{} is based on a Generative Adversarial Neural Network (GAN). Our key insight is to use an image-to-image style transfer approach for our attack. \agan{} takes as input a biomedical image of a victim, and translates it to produce a new (fake) image that changes the ``style'' by injecting an attacker-chosen disease into the image, while preserving the ``content'' or identity of the victim. The targeted condition does not reflect the real health condition of the patient. Therefore, the key outcome is to produce a ``deepfake'' image that can cause both a human (medical professional), and an algorithm to misdiagnose the health condition~\cite{rajpurkar2017chexnet, ai_health_news4, Jo_2019}. This is because the produced image shows human-perceptible signs of the targeted condition, which is also sufficient to mislead automated algorithmic schemes.

An incorrect diagnosis can lead to potential life-threatening situations for the patient, unnecessary healthcare costs, and wasted healthcare resources. The attacker may be motivated by financial gain. For example, a malicious clinic can trigger misdiagnosis and force an insurance provider to pay for unnecessary procedures. The attacker may also be motivated by criminal intent to cause harm to an individual (or group of individuals), or to disrupt and damage a particular healthcare framework~\cite{rushanan-sok2014, mirsky2019ctgan}.

A key aspect of \agan{} is the capability to perform \textit{controlled} creation of the fake image. Apart from being able to inject an attacker chosen disease condition, it can be done while preserving the ``identity'' of the patient. This is important because biomedical images are known to contain patterns that are unique to their patients. Preserving identity makes the attack much harder to detect because the image will appear to belong to the victim and will not seem abnormal. To make the attack more damaging, our tool can also be used during repeat visits by the victim to inject disease patterns that mimic the natural progression of the disease.

Attacks using \agan{} are realized by training on publicly available medical image datasets. Our attack only requires datasets annotated with health conditions and (anonymized) patient IDs. More importantly, \agan{} only requires a single image of the victim to generate a misleading \textit{fake} image.

Key contributions of our work include the following:\\
(1) We design and implement a GAN-based tool called \agan{} that can inject an attacker chosen disease condition into a victim's image, while preserving their identity. \\
(2) We demonstrate the feasibility of the attack on two popular biomedical image modalities---X-rays and retinal fundus images. Using publicly available medical datasets, \agan{} is used to inject Cardiomegaly, and Pleural Effusion health conditions into chest X-rays of healthy patients. For retinal images, we demonstrate injection of the Diabetic Retinopathy condition.\\
(3) We show that attacks powered by \agan{} can be sustained over time. As patients make repeat visits to a hospital, \agan{} can be used to inject disease conditions that match the expected progression of a disease.\\
(4) The effectiveness of attacks by \agan{} is evaluated by both (a) machine learning algorithms and image quality metrics, and (b) medical professionals. Our user study shows that medical professionals are convinced of the presence of targeted disease conditions in the fake images, and that they are unable to distinguish between real and fake images.\\
(5) Finally, we explore defensive schemes. We investigate two machine learning-based detection schemes: (1) \textit{Blind detection}: assumes no access to fake images, and no knowledge of attacker's model for training, and (2) \textit{Supervised detection}: assumes access to both real and fake images for training. We show that supervised detection schemes are highly effective, but also vulnerable to evasion schemes that modify \agan{} to bypass detection.

\section{Background and Related Work}\label{sec: background}

\para{Problem motivation.}
Healthcare spending is huge in many developed countries. In 2017, the US spent 17\% of its GDP on healthcare~\cite{ushealthspending18}. Not surprisingly, given the money involved, the healthcare system is already fraught with fraud~\cite{finlayson2019adversarial, al2018fraud, morris2009combating, rashidian2012no, bauder2017medicare, herland2018big}. Different entities/actors in the system---medical institutions (large hospitals as well as small clinics), medical practitioners (e.g., physicians, radiologists), health insurers, all have an incentive to engage in fraud and benefit financially~\cite{finlayson2019adversarial}. For example, a clinic can bill patients for unnecessary procedures or medication~\cite{wennberg2007extending}. On one hand, hospitals are known to inflate the cost of medical care to overcharge patients, while on the other, insurers have an incentive to reduce payout~\cite{joudaki2015using, barlett2006critical}. The attacker may also be motivated by a criminal intent to cause harm to an individual (or group of individuals), or to disrupt and damage a particular healthcare framework~\cite{rushanan-sok2014}. Overall, security practices implemented in healthcare systems are lacking, making them vulnerable to attacks that compromise the availability and integrity of medical data~\cite{beek_2018, verizon_2018}. These trends motivate us to explore potential threats by malicious actors that leverage technological advances in machine learning to engage in hard-to-detect medical fraud.

\para{Biomedical images.} 
Medical imaging is a crucial component of any health care framework. Biomedical images taken using specialized instruments capture interior anatomical structures of a human body. By analyzing these images, medical practitioners can monitor diseases and prepare treatment plans, often without requiring any invasive procedures~\cite{reddy2015healthcare}. To demonstrate our attack, we focus on two widely used image modalities, X-rays, and retinal fundus images. 
X-ray images are widely used to assess a range of injuries (e.g., damaged bones), and health conditions such as heart disease, breast cancer, and collapsed lungs.
Fundus photography helps to capture an image of the back of the eye. Fundus photographs of the retina~\cite{gulshan2016development} are used by ophthalmologists to detect Diabetic Retinopathy, a condition that could lead to vision loss in patients with diabetic mellitus~\cite{gulshan2016development}.

Attacks proposed in this work, while demonstrated on X-ray and retinal fundus imagery, are theoretically applicable to other 2D image modalities, \eg Ultrasound, and 2D MRI scans~\cite{reddy2015healthcare}. In Section~\ref{sec:threat-model}, we also explain potential application to 3D modalities. 
Also, unlike other images, biomedical images are usually highly standardized in terms of anatomical position and exposure~\cite{simon1972principles}. This makes it easier to learn patterns in anatomical structures and makes them more vulnerable to attacks discussed in this work.

\para{Adoption of ML for healthcare decisions.} Given the availability of large medical image datasets, and advances in deep learning, it is expected that algorithms will play a significant role in aiding healthcare decisions~\cite{gulshan2016development}. Deep learning schemes that analyze medical images can help doctors spot health conditions that may be otherwise hard to identify even by a trained professional~\cite{rajpurkar2017chexnet}. Insurance providers may also leverage algorithms to automate verification of diagnoses, before making reimbursements~\cite{finlayson2019adversarial}. Recently, the community has seen rapid advances in algorithmic decision making for medical imaging tasks that even surpass human performance. Recently, the U.S. Food and Drug Administration approved an AI algorithm to screen chest X-rays for collapsed lung (or Pneumothorax)~\cite{ucsf_xray_fda}. In fact, AI-based systems are already being tested/deployed to assess diabetic blindness~\cite{ai_health_news3}, detect chromosomal abnormalities~\cite{ai_health_news5}, and pancreatic cancer~\cite{ai_health_news6}. Recent examples also include ML schemes that perform well on breast cancer detection~\cite{wang2016deep}, skin cancer classification~\cite{esteva2017dermatologist}, arrhythmia detection~\cite{rajpurkar2017cardiologist}, hemorrhage identification~\cite{van2016fast}, and diabetic retinopathy detection~\cite{rajalakshmi2018automated}.

Hence, a successful attack that aims to mislead any medical decision process, should also fool an ML scheme designed for the same decision process. Otherwise, even if it can mislead a medical professional, it could be thwarted by an ML scheme.

\para{\agan{} vs attacks using adversarial samples.} Given an input image, one can add carefully crafted adversarial perturbations that are imperceptible to humans, such that the perturbed input triggers misclassification when fed to a model~\cite{ling2019deepsec}. Therefore, one approach to mislead ML-based  diagnostics tools is to craft adversarial samples of biomedical images---given a chest X-ray of a healthy patient, craft an adversarial input that fools the ML model to predict a disease condition~\cite{finlayson2019adversarial}. However, today, medical images still undergo visual examination by professionals---radiologists can identify that an adversarial X-ray image (targeting a disease condition) still looks healthy, thus rendering the attack ineffective. \textit{Hence, we propose attacks that produce images containing visually perceptible changes that can  mislead both a human (more importantly, a medical professional) and an ML model.}

\para{Prior work on misuse of ML/AI.} Most prior work at the intersection of ML and security primarily focused on attacks against ML systems~\cite{yuan2019adversarial}. But, recently, there is emerging interest in understanding new attacks enabled by ML/AI~\cite{brundage2018malicious}.
In the non-medical space, research has demonstrated AI techniques to break CAPTCHA systems~\cite{goodfellow2013multidigit}, generate convincing fake reviews~\cite{juuti2018stay, yao-2017-dnnreviews, radford2015unsupervised} and email content~\cite{das2019automated}, control voice assistants (e.g., Google Assistant, Alexa)~\cite{yuan2018commandersong, iter2017generating}, extract private information from collaborative learning systems~\cite{hitaj2017deep}, attack anonymity systems (Tor)~\cite{Nasr_2018, sun2015raptor}, and to automate DoS attacks via trace synthesis~\cite{yan2019automatically}.

In the medical domain, there is limited work on understanding threats posed by AI. Kohli et al. proposed an Iris presentation attack using a DCGAN~\cite{kohli2017synthetic}. Their attack specifically focused on Iris biometric systems and is not generally applicable.  In more closely related concurrent work from 2019, Mirsky et al. proposed CT-GAN~\cite{mirsky2019ctgan}, a framework that uses deep learning to tamper with 3D medical imagery (CT images) to add or remove signs of medical conditions. Compared to both these works, we propose a generic attack that is applicable to multiple diseases, multiple modalities (X-ray, retinal fundus), and requires significantly less effort from the attacker to successfully mislead diagnostic processes. We discuss CT-GAN in more detail in Section~\ref{sec:threat-model}.

\section{Attacking Medical Diagnostics}
\vspace{-1ex}
\subsection{Threat Model and Overview of Approach}
\label{sec:threat-model}
\vspace{-1ex}
\para{Threat Model.} 
The victim, a patient, visits a medical imaging lab, and obtains a biomedical image of some type, \eg X-ray, retinal Fundus. Unknown to the victim, the attacker obtains access to the victim's image, and ``translates'' it to a version with an attacker-chosen disease condition (that does not reflect the actual health situation of the victim). 
After translation, the original image is removed from the system by the attacker and is not seen by anyone else. For example, given an X-ray of a healthy patient, the attacker generates a new image indicating an abnormal heart condition when examined by either a medical professional or a machine learning algorithm. Figure~\ref{fig:attack-flowchart} illustrates the attack scenario. It is important to note that medical professionals will analyze the image for presence of any diseases. As the generated image is examined by a professional, traditional adversarial sample attacks, or na\"ive attacks that tamper with the diagnostic end-results are ineffective.
We make the following assumptions:

\begin{figure}[t]
    \centering
    \includegraphics[width=8.2cm]{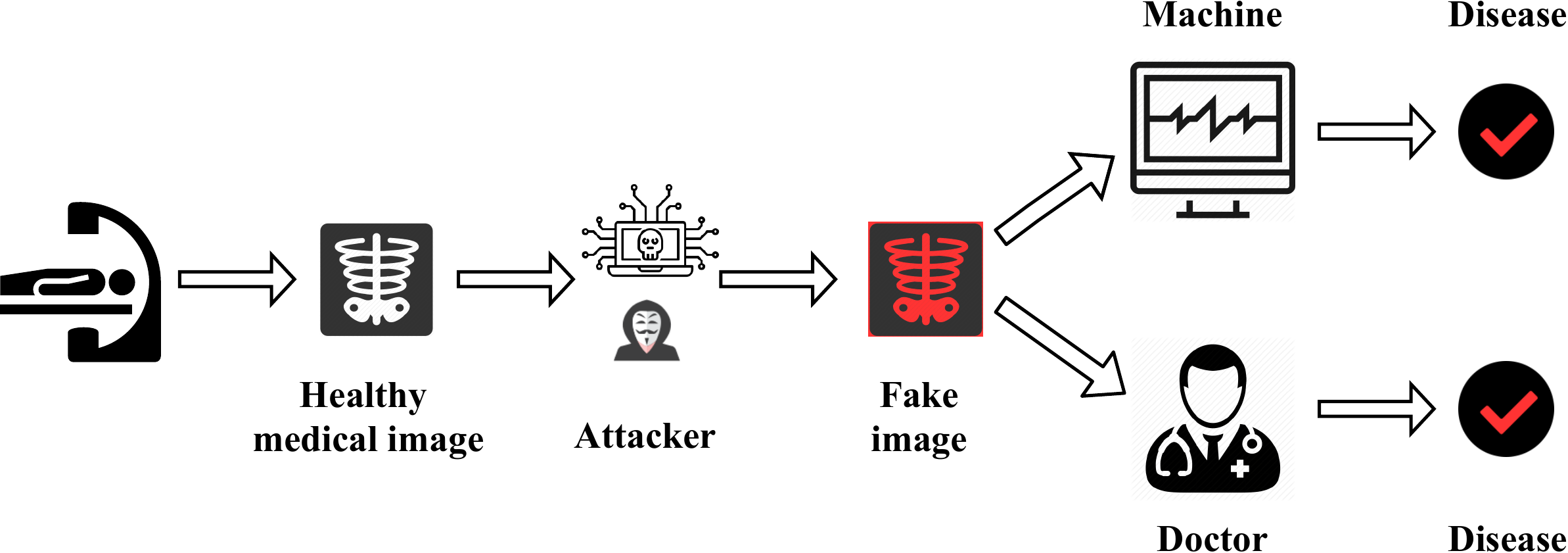}
    \vspace{-3mm}
    \caption{Overview of attack scenario.} 
    \vspace{-4mm}
    \label{fig:attack-flowchart}
\end{figure}

\textit{First,} the attacker can access and alter victim's medical images by compromising medical information systems. Medical images are typically managed through a Picture Archiving and Communication System (PACS)~\cite{pacs_overview}. The attacker can either access the data-at-rest, \eg when it is stored on a PACS server, or when data-is-in-motion, \eg by intercepting network traffic or by reading from volatile memory. Real world deployments of PACS are known to follow poor security practices, \eg misconfigurations, using default credentials, easy access by insiders, rare security patch updates, and lack of encryption support~\cite{verizon_2018, beek_2018}. These flaws cause the PACS systems to be highly vulnerable to social engineering and physical insider attacks, thereby violating data integrity. In fact, Mirsky et al.~\cite{mirsky2019ctgan} demonstrated a successful pen-test using a Raspberry Pi that was able to compromise a hospital's PACS framework, making it feasible to intercept scans and credentials of staff members in plaintext, and install a malware that allowed for a \textit{man-in-the-middle} attack. PACS servers are also exposed to and thus accessible via the internet~\cite{pacs_overview}. This exposure amplifies the susceptibility to attacks via web portals~\cite{beek_2018}.

\textit{Second,} the attacker can leverage publicly available medical image datasets to build ML models for the attack. There are large datasets covering different image modalities, e.g., X-ray~\cite{wang2017chestxray}, retinal fundus images~\cite{gulshan2016development}, MRI~\cite{oasis}, CT~\cite{yan2018deeplesion, armato2015data}, and PET~\cite{oasis}. The attack requires annotations that capture anonymized patient IDs, health conditions, \ie whether patient is healthy or has some disease(s), and any available characteristics such as the stage or severity of the condition. \textit{Note that the attack does not require any disease segmentation masks for these images, e.g., a marked portion of the image that shows an area affected by disease.}

\textit{Third,} the attacker has no access to any previous medical images of the victims. The attack only requires a single image from the victim, which can be accessed after the patient undergoes a medical imaging procedure. 

\textit{Lastly,} no knowledge is assumed regarding how the image is analyzed by the entity responsible for the victim's healthcare. Images could be analyzed by a medical professional or a machine learning algorithm (or both). Also, we do not assume any query access to the machine learning algorithm evaluating the victim's health condition. Such an access is not required in our attack because the attack visibly injects the targeted disease in the image, unlike  traditional adversarial inputs where imperceptible adversarial noise is added to fool classifiers~\cite{goodfellow2014explaining}.

\para{Attack goals.} Three main goals include the following:

\noindent \textit{1. Translate a biomedical image to a new one that indicates an attacker chosen health condition.} 
The targeted condition does not reflect the real health condition of the patient. 
In the rest of the paper, we use the phrase ``disease injection'' to refer to this image translation process, and the generated image is called the \textit{fake image}. While we primarily focus on injecting a disease condition, our methods can also be used to translate an image with a disease condition to one that appears healthy.

\noindent \textit{2. Inject disease while preserving identity of the victim. }For a successful attack, one must not only inject the disease, but must also ensure that the fake image reflects the ``identity'' of the victim. Biomedical images typically contain patterns that are unique to their owners, and such personal signatures can be used to verify the identity of the owners~\cite{qamber2012personal}. This applies to modalities such as X-rays, and retinal images that carry unique signatures of the patients. 
If the attacker used a diseased X-ray from another patient or generated a chest X-ray image that fails to preserve the anatomical characteristics of the victim, the image could be flagged by an identity verification algorithm, or by a doctor on visual examination. In such cases, the doctor or algorithm can compare the generated image against previously submitted images of the patient to verify the identity, thus rendering the attack ineffective.

\noindent \textit{3. Sustaining the attack over time.} Patients typically undergo repeated examinations to follow up on a health condition. This provides an opportunity for the attacker to continue to manipulate the system. To enable subsequent attacks, the attacker would need to control the disease injection process to reflect the natural progression of a health condition. Otherwise, there is a risk of raising suspicion and being caught. Such repeated attacks can be devastating for the patient and lead to wasted resources in the health-care system. For this goal, we propose methods to sustain the attack over time.

\para{Comparison to CT-GAN~\cite{mirsky2019ctgan}.} 
The only closely related work is CT-GAN, a deep-learning framework for 3D medical image tampering. The key differences between \agan{} and CT-GAN are as follows: 
\textit{(1) Practicality and efficiency.} To inject a chosen disease condition, CT-GAN requires at least 14 complex steps, including extensive pre-processing steps, identifying the region for disease injection, and multiple manual touch-ups to make the tampered region look realistic. \agan{} is more efficient to use, as it only requires the single translation step of passing the image through the generator. \agan{} is also more practical for a generalized bad actor, as it requires no knowledge of medically viable regions for disease injection. \textit{(2) Generalization to non-localized conditions.} CT-GAN uses an ``in-painting'' scheme to inject a disease into a specific image region. However, this not only requires that the chosen region be a medically viable location for the disease, but also requires the disease to be localized in the said region. \agan{}, on the other hand, learns to inject the disease as a whole into the image. This enables injection of diseases that are spread unevenly over multiple anatomical regions \eg injecting diabetic retinopathy into retinal fundus images. It is unclear if CT-GAN can work with such modalities. \textit{(3) Image Dimensionality.} CT-GAN is intended for use with 3D medical image modalities \eg CT scans. Disease injection is performed by extracting a 2D image slice from the middle of 3D DICOM imagery for the CT scan, applying the in-painting technique, and re-inserting the modified slice. \agan{} performance is demonstrated on 2D medical imagery but can potentially be extended to 3D modalities by performing single-step translation on a similarly extracted middle slice. 
Overall, \agan{} and CT-GAN demonstrate new threats facing our healthcare system.

\vspace{-1ex}
\subsection{Attack Methodology}
\label{sec:methodology}
\vspace{-1ex}
\noindent We use a Generative Adversarial Network (GAN)~\cite{goodfellow2014generative} for the attack. A GAN has two primary components, a \textit{generator} and a \textit{discriminator} that are trained in an adversarial process. Given a dataset of images, the generator 
learns to generate synthetic images that mimic the distribution of the dataset. The discriminator learns to decide whether an image produced by the generator looks real (i.e., belongs to the true data distribution) or fake. 
The two components are trained using a minimax objective where the generator aims to produce fake images that are indistinguishable from real images, while the discriminator aims to rightly distinguish between real and fake images. GAN variants have shown impressive results for high quality image generation tasks~\cite{brock2018large, karras2017progressive, karras2018style}.

Prior work using GANs in the medical domain mainly focused on non-adversarial scenarios, \eg for data augmentation\cite{frid2018synthetic,shin2018medical,frid2018gan,madani2018chest}, de-identification~\cite{kohli2017synthetic}, anomaly detection in data~\cite{schlegl2017unsupervised}, feature extraction~\cite{son2017retinal,baumgartner2018visual}, and image segmentation~\cite{yi2018generative,Xue2018SegANAN,moeskops2017adversarial,yang2017automatic,li2017brain,jin2018ct}. Our goal is not to propose a new GAN model to advance the state-of-the-art in biomedical image generation. Instead, we show how a GAN-based approach can be used to launch attacks against medical diagnostics.

While many GANs have been proposed, not all of them are suitable for the attack. In a vanilla GAN, the generator takes as input a noise vector $z$ drawn from some distribution $p_{z}(z)$ (e.g., normal distribution) to generate an image, while the discriminator provides feedback to improve the generation process. Different input vectors will produce different images, but in general it is hard to reverse engineer how the noise space maps to specific semantic properties of generated images, e.g., to represent a disease or identity of a patient. Therefore, this does not fit our scenario---having \textit{only} a latent vector as input makes it hard to \textit{control} the generation process. Another challenge is that the attacker only has a single image of the victim. This rules out approaches that train a GAN on past images of the victim to generate identity preserving images. Instead, we propose to learn from medical images of other patients (publicly available data) for the attack.

\para{\agan{}: Our attack framework.} In this section, we present \agan{}, a GAN-based image style transfer model for attacking medical diagnostics.
A style transfer GAN that takes an input image as a \textit{condition}, and ``translates'' it to a version that preserves the \textit{content}, while changing the \textit{style}~\cite{isola2017image}. In our context, content includes image characteristics that capture identity of the patient, while style captures the health condition of the victim. Note that using an image as input provides more control over the generation process, compared to a vanilla GAN.

One challenge is that an image-to-image translation GAN requires paired input-output data. For example, Pix2Pix GAN~\cite{isola2017image} learns from paired data to transfer style. In our setting, this would require a pair of images belonging to each patient, i.e., one with no disease, and the other with a disease. Such paired data is usually not publicly available and can be challenging for the attacker to obtain. Instead, we propose to do style transfer from \textit{unpaired collections of images}, \eg a set of images of arbitrary patients with no disease, and another set with a disease (again an arbitrary set of patients). The recently proposed CycleGAN~\cite{zhu2017unpaired} best fits this scenario, and we propose to build on top of this approach. \textit{A key advantage of our approach is that we will automatically learn characteristics of the disease (style), and identity (content) from the image collections, without requiring any human intervention or image segmentation masks (that highlight regions indicative of disease or those that capture identity of a patient). This enables a single-step disease injection attack, unlike prior work (CT-GAN).}

\para{\agan{} design.} Let $X$ be the domain of images having health condition $C_X$, and $Y$ be the domain of images diagnosed with another condition $C_Y$. Training samples in $X$ are ${\{x_i\}}_{i=1}^{N}$, and in $Y$ are ${\{y_i\}}_{i=1}^{M}$. Our goal is to translate images from $X$ to $Y$, \ie inject a new disease condition while preserving the identity of the patient. Without loss of generality we consider the victim to be a healthy person. We consider health condition $C_X$ to denote a healthy condition, i.e., with no disease, and $C_Y$ to denote a \textit{single} real disease. \agan{} can also be used for disease removal, in which case $C_X$ would include patients with a disease, and $C_Y$ will be healthy patients. %

An optimal image translator can translate images in $X$ to ones that match the distribution of images in $Y$. But because of the unpaired data setting, there are infinite possible mappings from $X$ to $Y$, and we want to produce a specific mapping---one that preserves the identity, while injecting a disease. We will later explain that this will require the use of two generators, and two discriminators to enable transfer from one domain to the other. One generator, $G:X\rightarrow Y$, transfers images from $X$ to $Y$, and another generator $F:Y\rightarrow X$ transfers images from $Y$ to $X$. One discriminator $D_{Y}$ tries to distinguish real images in $Y$ from images generated by $G$, another discriminator $D_{X}$ tries to distinguish real images in $X$ from images generated by $F$. 
Below, we explain how \agan{} enables style transfer for the attack. \agan{} is trained to optimize an objective function that includes the following loss terms.

\textit{Adversarial process.} This part captures the basic \textit{adversarial loss} for the GAN. A generator produces images that fall in a certain domain, and the discriminator tries to differentiate between generated and real images. We use a least-squares GAN loss~\cite{zhu2017unpaired} for our adversarial loss  to stabilize training and to improve image quality. It is computed as:
\begin{equation}
    \label{eqn:adv_loss_basic}
    \begin{split}
        L_{GAN}(G,D_{Y},X,Y) &=  \mathbb{E}_{y \sim p_{data(y)}}[D_{Y}(y)^2] \\ &+ \mathbb{E}_{x \sim p_{data(x)}}[(D_{Y}(G(x))-1)^2]
    \end{split}
\end{equation}
We use a similar adversarial loss function for the reverse direction, $F: Y \rightarrow X$ using the discriminator $D_X$. The final adversarial loss is as follows: $ L_{adv} =  L_{GAN}(G,D_{Y},X,Y) +  L_{GAN}(F,D_{X},X,Y)$.
The two generators try to minimize this objective, while the two discriminators try to maximize it.

\textit{Disease Injection.} In theory, the adversarial loss should be sufficient to inject a disease. However, we empirically observe that this is not the case. Disease characteristics in biomedical images can be subtle, e.g., small change in heart shape in a Chest X-ray, minor changes in a retinal vascular pattern. It is simple for the GAN to trivialize the style differences and simply replicate the input image. Thus, we incorporate an additional \textit{disease loss} term to enforce disease injection. 

A pre-trained disease classifier $S$ is used to calculate the disease loss and provide additional feedback to the generator. Generated images from $G$ are fed into $S$ to obtain a prediction probability $S(G(x)|C_Y)$ for belonging to disease condition $C_Y$. If $G(x)$ receives a high prediction probability to be in class $C_Y$, we add a small penalty, and a high penalty if the prediction probability is low. This pushes \agan{} to correctly inject the targeted disease condition. We define
\begin{equation}\label{eq: loss_disease}
    \begin{split}
        L_{disease} = \mathbb{E}_{x \sim p_{data(x)}}[l(S(G(x)|C_Y),C_Y)]
    \end{split}
\end{equation}
where $l$ is a cross entropy function.

\textit{Preserving identity.} As these are unpaired images, the adversarial loss term can map one image in $X$ to any random point in domain $Y$. This is not desirable, because we want to find a mapping that preserves identity. To reduce the space of possible mappings in $Y$, we draw on work from CycleGAN and apply a \textit{cycle consistency loss} to the GAN. 
Put simply, an image $x$ when translated to $Y$ and reconstructed back to domain $X$ should be mostly similar to the original $x$, i.e., $F(G(x))\approx x$. 
Similarly, there is a reverse cycle loss for translations from $Y$ to $X$ as well. The cycle loss is computed as:
\begin{equation}
    \begin{split}
        L_{cycle} &= \mathbb{E}_{x \sim p_{data(x)}}[\Vert F(G(x))-x\Vert_{1}] \\ &+ \mathbb{E}_{y \sim p_{data(y)}}[\Vert G(F(y))-y\Vert_{1}]
    \end{split}
\end{equation}
While we expect the cycle loss to find mappings to the other domain that are easier to reconstruct, it still lacks a concrete notion of identity. This is because the cycle loss formulation does not explicitly characterize what defines the identity of the patient. We argue that cycle loss is not sufficient to preserve identity all the time.  To better preserve identity, we propose an additional \textit{identity loss} term defined as perceptual loss given by
\begin{equation}
    \begin{split}
        L_{identity} &= \mathbb{E}_{x\in p_{data}(x)}[\Vert E(x)-E(G(x))\Vert_{1} \\ &+ 
        \Vert E(F(G(x)))-E(G(x))\Vert_{1})]
    \end{split}
\end{equation}
where $E(.)$ represents features extracted from a specific layer in a pre-trained identity classifier. Recall that attacker has a single image of the victim, so it is hard to train an identity classifier that includes all victims. \textit{Using a perceptual loss as opposed to a classification loss (as in the disease loss term) helps to overcome this issue. Perceptual loss allows us to use an identity classifier trained on any available set of patients because we only use features from an internal layer.}

Finally, the overall loss is computed as:
\begin{equation}
    \begin{split}
        L(G,F,D_{X},D_{Y}) &= \lambda_{adv}L_{adv} + \lambda_{disease}L_{disease} \\ &+ \lambda_{identity}L_{identity} + \lambda_{cycle}L_{cycle}
    \end{split}
\end{equation}
and the associated weight terms control the extent to which each property is enforced.

\textit{Sustaining the attack over time.} Ideally, when victims undergo repeated examinations, disease injections should match the expected progression of a disease, \eg disease becoming severe over time. We present two ways in which \agan{} can be used to enable such repeated attacks: \\
(1) Attacker can use publicly available datasets that capture different stages of the disease in question, and create multiple \agan{} models, each one trained to inject a specific stage of the disease into the patient's image. In fact, such datasets exist---we use a dataset of retinal fundus images to inject different stages of Diabetic Retinopathy (Section~\ref{sec: sustaining_attack_over_time}).\\
(2) If there is no data capturing progression of a disease, then we propose a simple alternative solution. Given a dataset capturing a certain (late) stage of a disease, and a healthy stage, attacker can inject intermediate stages of the disease using simple \textit{linear interpolation} over the available images. More specifically, the attacker will train a single \agan{} model to translate a non-disease image to a disease stage (for which data is available). Next, given a non-disease image $I_{nd}$ belonging to a victim, and a disease injected image, $I_{d}$ produced by \agan{}, attacker can use linear interpolation to approximate intermediate stages (represented by $I_f$) of the targeted disease as follows: $I_{f} = \alpha \cdot I_{d} + (1 - \alpha) \cdot I_{nd}$. Here $\alpha$ represents the degree of disease injection. Such injection is possible because \agan{} produces output images that are perfectly aligned with input images. In Section~\ref{sec: sustaining_attack_over_time}, we show how one can produce convincing attack images capturing intermediate stages of Cardiomegaly (heart condition) using this approach. However, we acknowledge that such interpolation schemes may not be meaningful for all disease conditions.

\para{\agan{} model architecture.}\textit{Generator and Discriminator.} Architecture for both generator and discriminator is inspired by CycleGAN~\cite{zhu2017unpaired}, and we build on top of a publicly available implementation from GitHub~\cite{CycleGAN-tensorflow}. Input and output image resolutions of the generator are $256\times256$. 
Discriminator is based on a $70\times70$ PatchGAN~\cite{isola2017image}, that decides whether $70\times70$ overlapping image patches are real or fake, resulting in a $32\times32\times1$ dimensional output. This has been shown to outperform a discriminator that evaluates the entire image to determine whether it is real or fake. More details of the generator and discriminator architecture are in Table~\ref{tab: generator architecture} in Appendix~\ref{appendix: models_and_datasets}. 
Note that \agan{} can be adapted to produce higher resolution images by borrowing architectural elements from PGGAN~\cite{karras2017progressive}. PGGAN can produce high resolution images by starting from a low resolution version, and progressively increasing the size (layers) of the network. %

\textit{Disease and Identity Classifiers.} Recall that \agan{} requires pre-trained disease and identity classifiers to preserve identity while injecting disease. We use the same architecture for both classifiers, but they are trained differently depending on the dataset.\footnote{This is because we use different types of transfer learning schemes.} Disease classifier is a binary classifier predicting condition as non-disease and the targeted disease. Identity classifier is a multi-class classifier predicting the identity of a person. Both classifiers use the DenseNet-121 model architecture~\cite{huang2017densely}. For both models, we replace the last classification layer of DenseNet-121 with a dense layer of 256 neurons, a dropout layer with rate of 0.5, followed by a final classification layer (that fits our task). To compute the identity loss, we extract the output of the convolution layer before the last dense block in DenseNet-121. More details of \agan{}'s architecture are in Appendix~\ref{appendix: models_and_datasets}.

\para{Alternative architectures for \agan{}.} Our techniques behind \agan{} (to inject disease and preserve identity) can be applied to other image-to-image translation GANs as well. 
In Section~\ref{sec:alternative-arch}, we investigate attack effectiveness when using alternate architectures. We explain two other architectures below.

\textit{StarGAN.} StarGAN is an image-to-image translation model that improves over CycleGAN~\cite{choi2018stargan} by providing many-to-many domain translation capabilities. Unlike CycleGAN, StarGAN only requires a single generator and discriminator, but leverages an auxiliary domain classifier to ensure successful domain translation. StarGAN also includes a cycle loss term similar to CycleGAN to preserve content. We adapt StarGAN to fit into our \agan{} framework by adding a disease and identity loss term. To compute disease loss, we use the auxiliary domain classifier available in StarGAN, but use an external classifier for identity loss. Additionally, we observed that StarGAN's auxiliary classifier suffers significantly when classes are imbalanced. To deal with such imbalance, we upsampled the underrepresented class, and replaced the binary-cross entropy loss used for the domain classification with focal loss \cite{lin2017focal}. More details are in Appendix~\ref{appendix: models_and_datasets}.

\textit{IPCGAN}. Identity Preserving Conditional GAN or IPCGAN is an image translation GAN to synthesize face images in a targeted age group, while preserving identity, \eg translate a teenager's face image to one that looks 50+, while preserving identity. To achieve this, IPCGAN uses an age classifier to enforce translation to the new age group and implements identity loss as perceptual loss. However, IPCGAN lacks any kind of reconstruction/cycle loss, as used in \agan{}. To fit IPCGAN into \agan{} framework, we replace the age classifier by a disease classifier, and use our patient identity classifier. We do not add a cycle loss term to its training objective. More details are in Appendix~\ref{appendix: models_and_datasets}.

\vspace{-1ex}
\section{Experimental Setup for Evaluating Attack}
\vspace{-1ex}
\begin{table}[t]
    \caption{\# images used to train and evaluate \agan{}.}
    \begin{center}
        \def\arraystretch{1.2}
        \begin{tabular}{l|c|c|c}
            \hline
            \multirow{2}{*}{\textbf{Datasets}} & \multicolumn{2}{c|}{\textbf{\# Train images}} & \multirow{2}{*}{\textbf{\# Victim images}}\\
            \cline{2-3}
            & \textbf{Disease} & \textbf{Non-disease} & \\
            \hline
            \textbf{Cardiomegaly}  & 35,352\ &1,349\ &6,235\\\
            \textbf{Effusion}  & 35,352\ &4,977\ &6,883\\\
            \textbf{Severe DR}  & 32,728\ & 1,000\ &680\\\
            \textbf{Proliferative DR}  & 32,728\ & 982\ &703\\
            \hline
        \end{tabular}
    \end{center}
    \label{tab:agan_data}
    \vspace{-6mm}
\end{table}

\noindent To build \agan{}, we need to first train the disease and identity classifiers, followed by the GAN component (that uses the pre-trained disease and identity classifiers). We build two versions of each of the disease and identity classifiers. One version of classifiers are used to train the \agan{}, and are called the attack disease ($C_{a}^{d}$) and identity ($C_{a}^{i}$) classifiers. The other version of classifiers are used to evaluate the success of the attack, and are called the evaluation disease ($C_{e}^{d}$) and identity classifiers ($C_{e}^{i}$). Given images translated by \agan{}, the evaluation classifiers help in answering the following questions. (1) \textit{Did we successfully inject the disease?} (2) \textit{Did we preserve the identity?} We ensure that attack and evaluation classifiers are trained on datasets with no patient overlap. 

For all attacks, we start with a victim set of healthy patients (i.e., having no diseases), and evaluate attack success by injecting different diseases (or different stages of same disease).

\vspace{-1ex}
\subsection{Medical Datasets}
\vspace{-1ex}
\para{NIH chest X-ray dataset~\cite{wang2017chestxray}.\footnote{\url{https://nihcc.app.box.com/v/ChestXray-NIHCC}}} This is a publicly available dataset of 112,120 frontal chest X-ray images of 30,805 unique patients. Images are annotated with anonymized patient IDs, with labels indicating presence of one or more of 14 diseases. We demonstrate disease injection for two of these diseases, namely \textit{Cardiomegaly}, and \textit{Pleural Effusion}. Cardiomegaly causes an enlarged heart, usually the result of heart disease. Pleural Effusion is a condition that causes buildup of excess fluid around the lungs. These conditions are chosen because prior work demonstrated high detection accuracy for both using deep learning~\cite{rajpurkar2017chexnet}.  

We partition the dataset by patients into two subsets. One partition is used for training \agan{}, including the GAN component, and the two attack classifiers ($C_{a}^{d}$, and $C_{a}^{i}$). The second partition is used only for evaluation which includes the victim set, and data for training the evaluation classifiers ($C_{e}^{d}$ and $C_{e}^{i}$). The attack partition contains 24,000 patients, and the evaluation partition has 6,805 patients. Partitioning was performed in a manner that allows us to build reliable evaluation classifiers, \eg victim set should be large, and  include patients with at least 10 images, so we could build a high quality identity classifier for evaluation ($C_{e}^{i}$). More details of data preparation are available in Appendix~\ref{appendix: models_and_datasets}.

The statistics of the data used to train and evaluate \agan{} are shown in Table~\ref{tab:agan_data} (see rows for Cardiomegaly and Effusion). Once trained, \agan{} is tested on over 6,000 victim images for both disease conditions. We also make sure that the victim set only includes patients with non-disease images (which can be then injected with a disease). Victim set includes images that correctly pass the disease classification test by $C_e^{d}$ (as non-disease), and the identity classification test by $C_e^{i}$ (as  having the correct identity). This ensures that any effected style transfer is due to the success of \agan{}, and not due to misclassifications by the evaluation classifiers. Details of dataset (from attack and evaluation partition) used to train the attack and evaluation classifiers are in Tables~\ref{tab: disease_classifiers_data} and~\ref{tab: identity_classifiers_data} in Appendix~\ref{appendix: models_and_datasets}.

\para{Retinal Fundus images.} This is a publicly available dataset\footnote{\url{https://www.kaggle.com/c/diabetic-retinopathy-detection/data}} provided by EyePACS, a platform for retinopathy screening. It consists of pairs (left and right eye) of retinal fundoscopy images for  88,702 patients. Images are annotated with anonymized patient IDs, with labels indicating different stages of Diabetic Retinopathy (DR)---no disease, mild, moderate, severe, and proliferative. DR is a disease impacting blood vessels in the retina leading to possible vision loss in people with diabetes. We demonstrate injection of severe and proliferative DR stages. %

We prepare the dataset following a similar methodology as used for chest X-rays. Dataset statistics are shown in Table~\ref{tab:agan_data} (see rows for severe and proliferative DR). 
For both stages, our victim set includes over 600 images. Tables~\ref{tab: disease_classifiers_data} and~\ref{tab: identity_classifiers_data} in the Appendix show statistics of data used for the attack and evaluation (disease and identity) classifiers. More details are in Appendix~\ref{appendix: models_and_datasets}.

\vspace{-1ex}
\subsection{Training \agan{}, and evaluation classifiers}
\vspace{-1ex}
\noindent All models are implemented using Tensorflow v1.12.0 framework for Python.\footnote{Only exception is the StarGAN version of \agan{} which is implemented in PyTorch.} An NVIDIA Titan Xp GPU with 12GB RAM, on a host with Intel(R) Xeon(R) W-2135 CPU @ 3.70GHz and 64 GB RAM was used for training.

\begin{table}[t]
    \caption{Testing accuracies of the disease and identity classifiers used.}
    \begin{center}
        \def\arraystretch{1.2}
        \begin{tabular}{l|c|c|c|c}
        \hline
        \textbf{Datasets} &$\boldsymbol{C^d_a}$ & $\boldsymbol{C^d_e}$ & $\boldsymbol{C^i_a}$ & $\boldsymbol{C^i_e}$ \\
        \hline
        \textbf{Cardiomegaly} &84\% &80\% &98.2\% &96.6\%\\
        \textbf{Effusion} &87\% &80.7\% &98.2\% &96.6\%\\
        \textbf{Severe DR} &89.9\% &90.4\% &98.5\% &99.9\%\\
        \textbf{Proliferative DR} &87\% &87.1\% &98.5\% &99.9\%\\
        \hline
        \end{tabular}
    \end{center}
\label{tab:classifier_accuracies}
 \vspace{-4mm}
\end{table}

\para{\agan{}.} For all experiments, the Adam optimizer is used with learning rate of 0.0002, $\beta_1=0.5$ and $\beta_2=0.999$. The learning rate remains unchanged for the first 100 epochs and is then decreased linearly for the next 100 epochs. For each dataset, we empirically determine the weights for each of the loss terms. Using a validation set, we empirically estimate weights that produce the highest quality images, while ensuring successful injection of disease and identity preservation. Training one instance of \agan{} takes $\approx$17 hours. Training configuration for each dataset and alternative architectures are in Appendix ~\ref{appendix: models_and_datasets}.

\para{Disease classifiers.} 
For both datasets (X-ray and retinal), we leverage transfer learning. For the X-ray datasets, the teacher model is trained on relevant partitions (attack or evaluation depending on the classifier) of the NIH Chest X-ray dataset to diagnose all 14 available diseases (multi-label classifier), using the training setup used by Rajpurkar et al.~\cite{rajpurkar2017chexnet}. To build our X-ray student model, we initialize our architecture (see earlier Section~\ref{sec:methodology}) with weights from the teacher model, and only fine-tune the last 70 layers. For the retinal DR disease classifiers, the teacher model is a DenseNet-121 architecture trained on ImageNet~\cite{huang2017densely}, and all layers are fine-tuned during training. Table~\ref{tab:classifier_accuracies} shows the accuracies of the (attack and evaluation) disease classifiers ($C^d_a$, and $C^d_e$) when applied to balanced test datasets. All classifiers have fairly high accuracy. Training configuration is available in Appendix~\ref{appendix: models_and_datasets}.

\para{Identity classifiers.} We again leverage transfer learning. For all datasets, the teacher model is a DenseNet-121 model trained on ImageNet. For each model, weights are initialized from the teacher model, and all layers are fine-tuned during training. The attack identity classifier, $C_a^i$, is trained to predict a random subset of patients in the \agan{} training dataset. It is not necessary to train the identity classifier on all patients in \agan{} training data, as we use a perceptual loss. For the retinal dataset, we perform data augmentation for both its training and testing data as we have limited data (only $2$ images per patient). Blurring and random rotations are used to augment the dataset and create a set of $14$ (including the original) images per patient. Table~\ref{tab:classifier_accuracies} shows the testing accuracies of the (attack and evaluation) identity classifiers (on balanced test datasets). All identity classifiers achieve over 96\% accuracy. Training configuration is in Appendix~\ref{appendix: models_and_datasets}.

\section{Evaluating Effectiveness of Attacks}
\vspace{-1ex}

\noindent We structure the evaluation of \agan{} based on our primary goals. More specifically, we aim to demonstrate that images generated by \agan{} show signs of disease and preserve patient identity. This requires misleading both real-life medical professional diagnostics, \textit{as well as} machine learning classifiers that are used to aid diagnosticians. Therefore, we perform evaluation by: (1) different machine learning tools, and image quality metrics, and (2) by consulting medical professionals.

\begin{figure}[t]
    \centering
    \includegraphics[width=0.95\columnwidth]{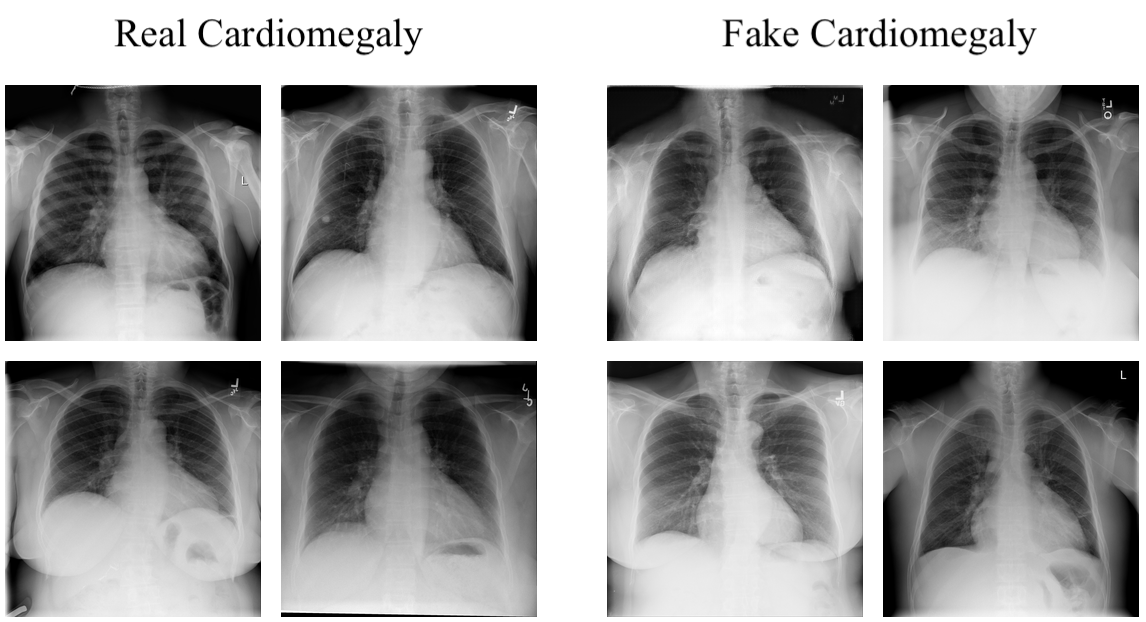}
    \caption{Real and fake X-ray images with Cardiomegaly.}
    \label{fig:cardio_real_fake}
    \vspace{-0.1in}
\end{figure}

\begin{figure}[t]
    \centering
    \includegraphics[width=0.95\columnwidth]{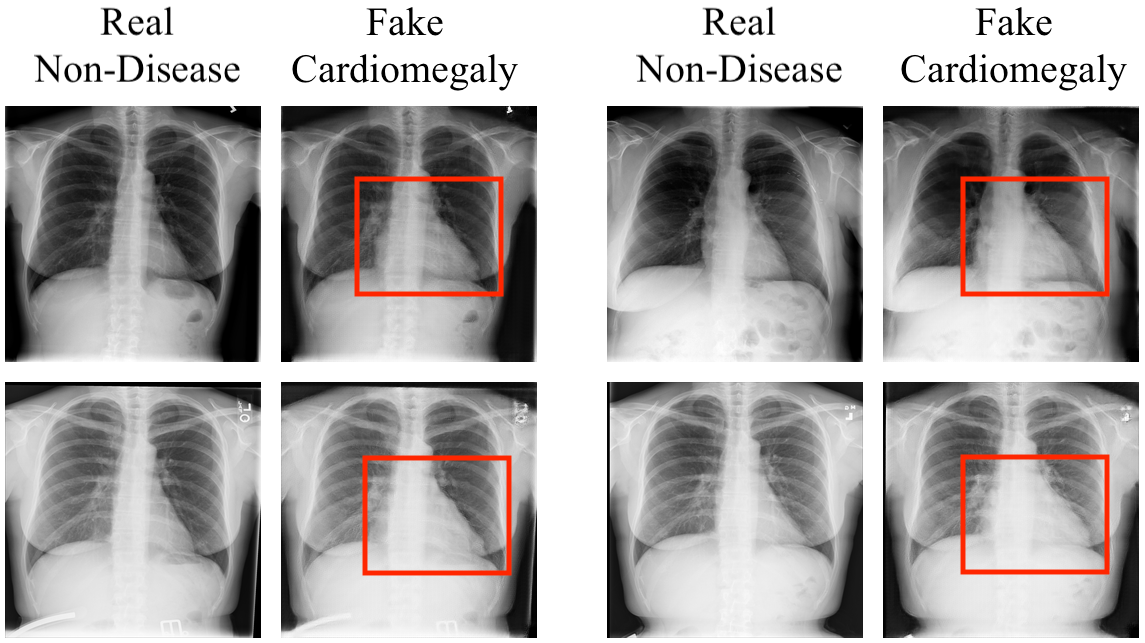}
    \caption{Injection of Cardiomegaly, red rectangles label the enlarged heart.}
    \label{fig:injection_cardio}
    \vspace{-0.2in}
\end{figure}

\subsection{Evaluation by Machine Learning Tools and Image Quality Metrics}

\noindent In this section, we examine different aspects of \agan's effectiveness and design in detail. This includes evaluating: (1) image quality, (2) disease injection, (3) identity preservation, (4) feasibility of sustaining the attack over time to match progression of disease, (5) effectiveness when using alternative architectures, and (6) design of \agan{}'s training objectives through an ablation study. For most of our evaluation, we focus on Cardiomegaly, and Effusion for X-rays, and proliferative DR for retinal images. We consider an earlier stage of DR---Severe DR, when evaluating progressive injection of disease.

\begin{figure}[t]
    \centering
    \includegraphics[width=0.95\columnwidth]{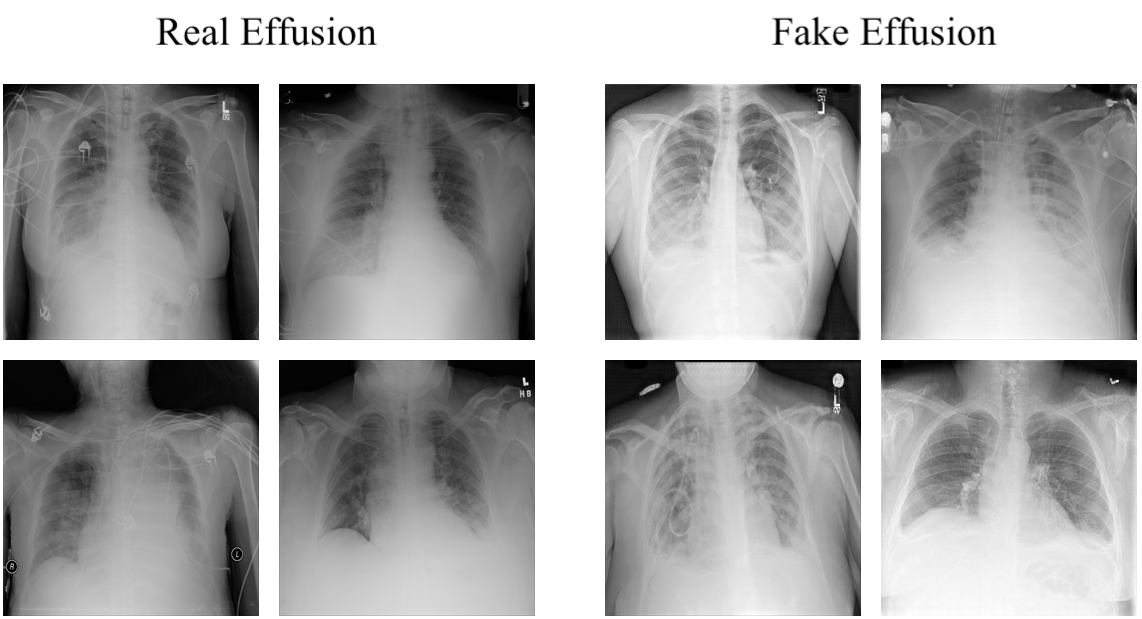}
    \caption{Real and fake X-ray images with Pleural Effusion.}
    \label{fig:effusion_real_fake}
\end{figure}

\begin{figure}[t]
    \centering
    \includegraphics[width=0.95\columnwidth]{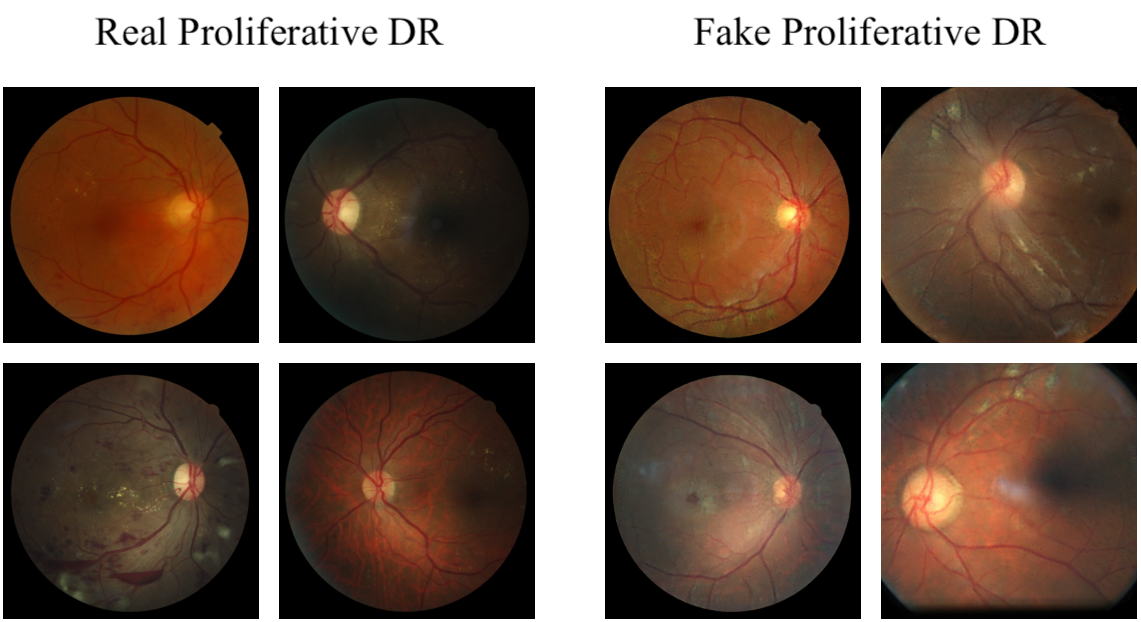}
    \caption{Real and fake Retinal Fundus images with Proliferative DR.}
    \label{fig:proliferative_real_fake}
    \vspace{-4mm}
\end{figure}

\begin{figure}[t]
    \centering
    \includegraphics[width=0.95\columnwidth]{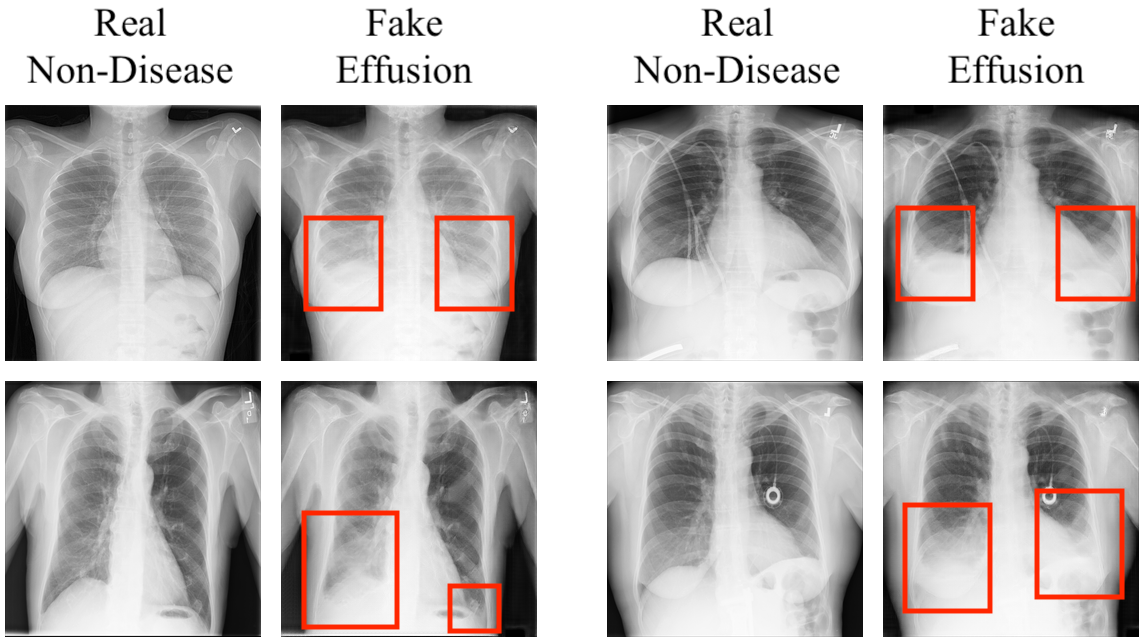}
    \caption{Injection of Pleural Effusion, red rectangles label the fluid.}
    \label{fig:injection_effusion}
    \vspace{-2mm}
\end{figure}

\vspace{-1ex}
\subsubsection{Image Quality}\hfill\\
We start with analyzing quality of images generated by \agan{}. Image quality is important, otherwise the fake images can be easily caught by human inspection.

\para{MSSIM.} We use Multi-scale Structural Similarity (MSSIM) metric to evaluate image quality. MSSIM is a widely used objective image quality metric that correlates well with perceived image quality~\cite{borji2018pros}. The metric is based on the idea that humans are sensitive to changes in structure, luminance, and contrast. This metric also considers perceived distortion from different viewing angles by analyzing the images at different scales. Given a reference image $X$, and a distorted version $Y$, $MSSIM(X,Y)$ ranges between 0 and 1, with 1 indicating that images are identical, and 0 indicating no structural similarity. 

For each dataset, we first compute the MSSIM score between random pairs of real disease images, say between the sets Real A and Real B. Next, we compute MSSIM between images in real A and a random set of fake images with the same disease type/stage. We expect the MSSIM scores between real and fake images to be similar or higher than the scores between real images. If the MSSIM score between real and fake images is significantly lower than between real images, it would indicate that fake images look very different from real images of the same type (i.e., disease). Table~\ref{tab: mssim_scores} shows the average MSSIM scores, and the scores are within a similar range, suggesting that perceived image quality of fake images when compared to real images is satisfactory.

Figures~\ref{fig:cardio_real_fake}, \ref{fig:effusion_real_fake} and \ref{fig:proliferative_real_fake} show samples of fake and real images with Cardiomegaly, Effusion, and Proliferative DR, respectively. In these figures, for each fake image, a real image which is closest to the fake image in terms of L2 score is chosen. Showing such similar samples of real images helps to better understand image quality. Overall, we find that \agan{} is able to generate high quality images that are hard to distinguish from real images. More image samples can be found in Figure~\ref{fig:appendix_real_fake} in the Appendix.

\para{FID and Inception score} Inception score~\cite{borji2018pros} and Fr\'echet Inception Distance~\cite{borji2018pros} are popular quantitative metrics to evaluate quality of GAN generated images. Both rely on extracting embeddings from models trained on the ImageNet dataset. However, they are not suitable in our case, because the ImageNet dataset does not include biomedical images.

\begin{figure}[t]
    \centering
    \includegraphics[width=0.95\columnwidth]{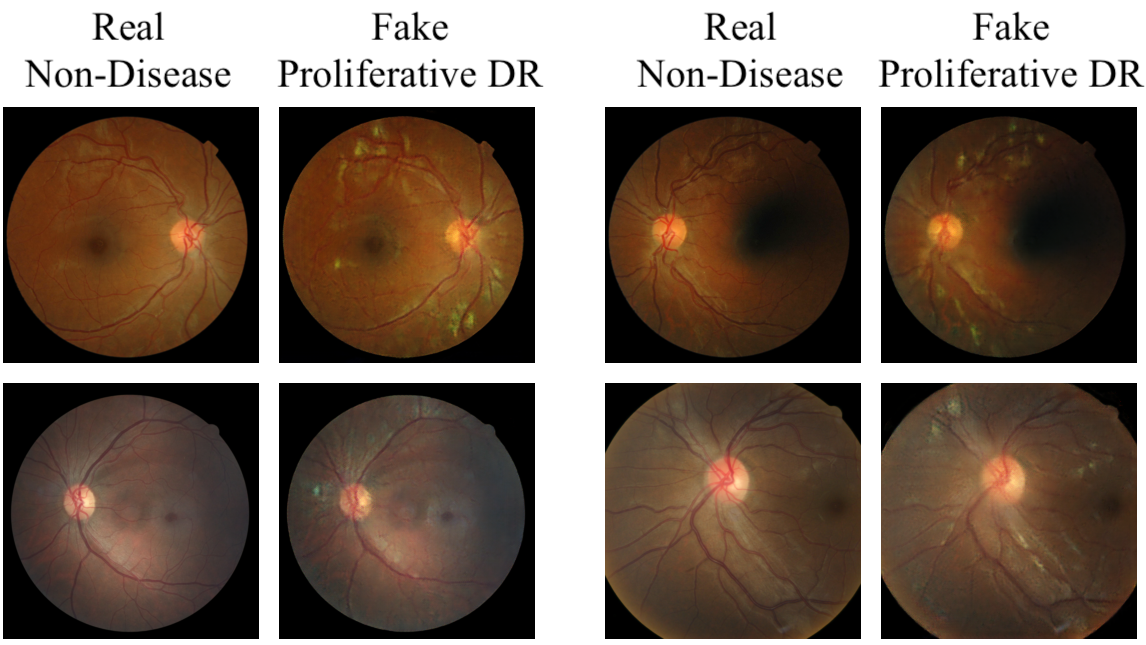}
    \caption{Injection of Proliferative DR. Cotton wool spots indicate disease.}
    \label{fig:injection_proliferative}
    \vspace{-4mm}
\end{figure}

\begin{figure}[t]
    \centering
    \includegraphics[width=1\columnwidth]{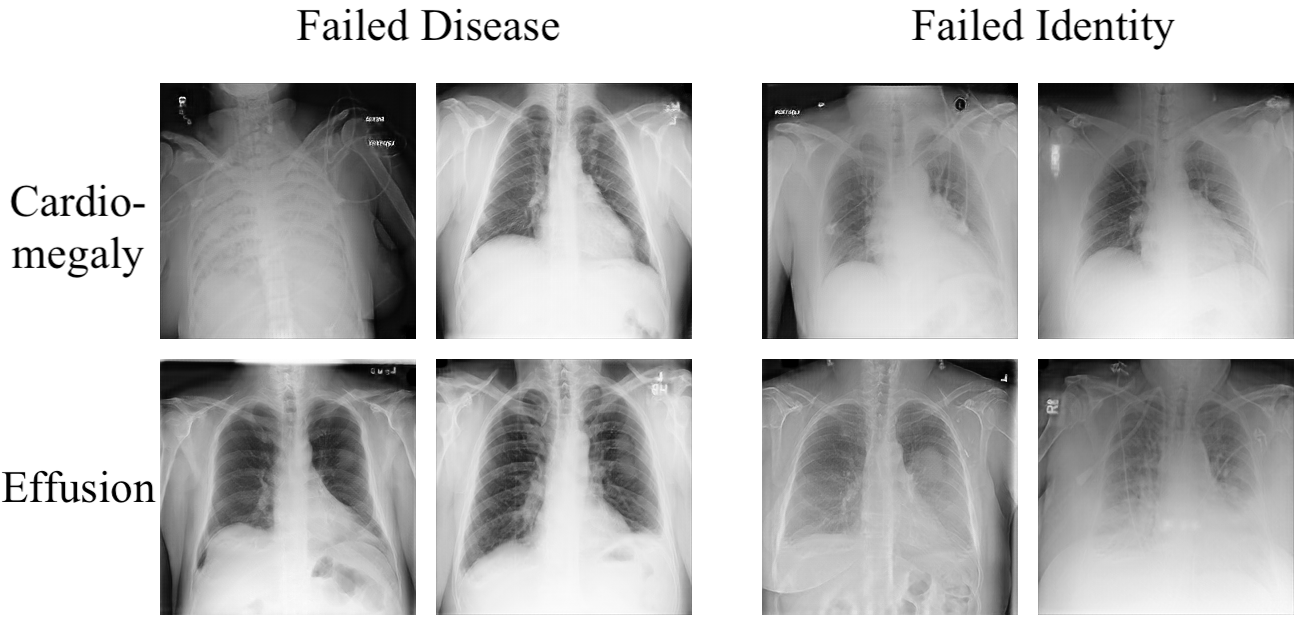}
    \vspace{-0.1in}
    \caption{Examples of failure cases for Cardiomegaly and Effusion.}
    \label{fig:xray_fail}
    \vspace{-2mm}
\end{figure}

\begin{table}[t]
    \caption{MSSIM scores to evaluate image quality of fake images.}
    \vspace{-2mm}
    \def\arraystretch{1.2}
    \begin{center}
        \begin{tabular}{l|c|c}
        \hline
        \multirow{2}{*}{\textbf{Datasets}} & \multicolumn{2}{c}{\textbf{MSSIM Score}} \\ \cline{2-3}
        & \multicolumn{1}{l|}{\textbf{Real A vs. Real B}} & \multicolumn{1}{l}{\textbf{Real A vs. Fake B}}  \\
        \hline
        \textbf{Cardiomegaly}  & 0.456 &0.449\\
        \textbf{Effusion}  & 0.439 &0.429\\
        \textbf{Proliferative DR}  & 0.516 & 0.515\\
        \hline
        \end{tabular}
    \end{center}
    \label{tab: mssim_scores}
    \vspace{-4mm}
\end{table}

\vspace{-1ex}
\subsubsection{Disease Injection Success}\hfill\\
\para{Disease injection rate.} Success of disease injection is measured using the evaluation disease classifier, $C_e^d$. We compute a \textit{disease injection rate}, $R_d$, as the percentage of generated images classified by $C^d_e$ as having the disease. The rates are listed in the second column of Table~\ref{tab: jekyll_results}. Disease injection rates are high for all diseases, with Proliferative DR showing 100\% injection rate.

Figures~\ref{fig:injection_cardio},~\ref{fig:injection_effusion}, and~\ref{fig:injection_proliferative} show successful disease injections of Cardiomegaly, Effusion and Proliferative DR, respectively. The real non-disease image of the victim is shown next to each fake image. Changes can be observed in the fake images compared to the non-disease images. For Cardiomegaly in Figure~\ref{fig:injection_cardio}, we see an enlarged heart (highlighted by red rectangle) compared to the non-disease image of the victim. \agan{} intelligently learned to selectively modify the heart region, while keeping rest of the image mostly similar. In Figure~\ref{fig:injection_effusion}, one can observe build-up of excess fluid outside the lungs (marked by the red rectangles) as expected for Effusion. In the case of retinal images in Figure~\ref{fig:injection_proliferative}, there are some noticeable changes as well---\agan{} learned to generate ``cotton wool''-like spots on the retina when DR is observed. This matches real description of symptoms for a patient with DR~\cite{AOA}. We provide more samples of disease injection in Figure~\ref{fig:appendix_injection} of the Appendix. %

To better understand why we failed in some cases (for Cardiomegaly and Effusion), we manually went through some of the failed examples. Failed examples of Cardiomegaly are shown in Figure~\ref{fig:xray_fail}. We are able to identify two types of failures: \textit{First}, there are cases where the generated image does show an enlarged heart, but maybe not large enough to pass the classifier test. \textit{Second}, the original victim images  did not have a distinct heart shaped region in the image, so it is harder for \agan{} to make modifications to that region. Failed examples of Pleural Effusion are shown in Figure~\ref{fig:xray_fail}. In failed examples, we observe only a \textit{partial} buildup of fluid on one or both sides of lungs, which, while visible, is not significant enough to pass the classifier test. More failed samples are available in the Appendix (Figures~\ref{fig:appendix_fail_cardio},~\ref{fig:appendix_fail_effusion}).

\begin{figure}[t]
    \centering
    \includegraphics[width=0.90\columnwidth]{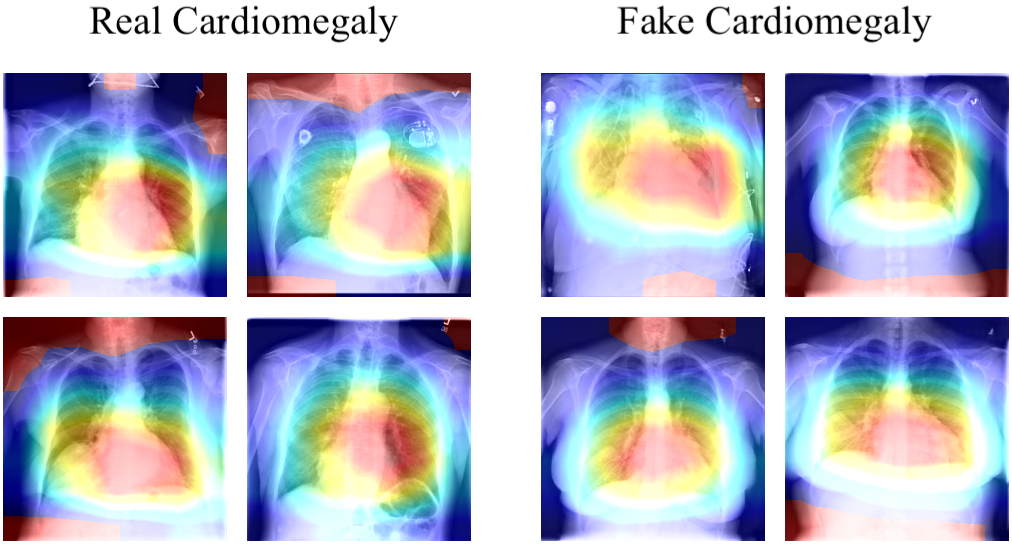}
    \caption{Examples of heat-maps of real and fake cardiomegaly X-rays.}
    \label{fig:heatmap_cardio}
    \vspace{-4mm}
\end{figure}

\para{Disease classifier model interpretation.}
To further evaluate disease injection success, we leverage prior work on chest X-rays to identifying regions most indicative of a disease. Put simply, we check whether a machine learning model thinks we injected the disease in the right way, e.g., is the condition injected in the right place? Rajpurkar et al.~\cite{rajpurkar2017chexnet} designed a model called ChexNet and present a method to identify regions of interest (for disease classification) in Chest X-rays by analyzing convolution feature maps. We use their tool and Figure~\ref{fig:heatmap_cardio} shows heat-maps visualizing regions indicative of Cardiomegaly (red means higher probability) for both real and fake images. In both real and fake X-ray images with Cardiomegaly, we observe red regions highlighting the central and upper-right regions of the heart. In these regions, we observe visual enlargement of the heart, which is indicative of Cardiomegaly. %

\vspace{-1ex}
\subsubsection{Identity Preservation Success}\hfill\\
\para{Identity preservation rate.} Similar to disease injection rate, we compute identity preserving rate, $R_i$ using $C_e^i$. The percentage of generated images classified into the correct patient ID is the identity preserving rate. This is listed in Table~\ref{tab: jekyll_results}, and is over 88\% for all datasets, with X-ray modalities showing over 94\% identity preservation rate. In Figures~\ref{fig:injection_cardio},~\ref{fig:injection_effusion}, and~\ref{fig:injection_proliferative}, characteristics of generated images are quite similar to the victim's original image, except for any changes due to disease injection. The generated X-ray images presented in Figures~\ref{fig:injection_cardio},~\ref{fig:injection_effusion} indicate preservation of anatomical structure (positioning and shape of ribs, thoracic cavity, lungs, etc.), while generated retinal images in Figure~\ref{fig:injection_proliferative} mostly preserve the vessel patterns as well.

We further investigated the samples that fail to preserve identity. Figures~\ref{fig:xray_fail} shows failed examples of Cardiomegaly, and Effusion. Failed examples of Proliferative DR are in the Appendix (Figure~\ref{fig:failed_proliferative_retinal}).
Overall, we suspect that injection of disease patterns disrupted other regions of the image that may have been crucial to capturing identity. For example, Effusion often introduces a fluid that can obscure large areas of the lung regions, hiding thoracic structure we believe to be crucial to patient identity. Similarly, in Proliferative DR, the `cotton-wool' spots that arise due to insufficient blood supply, can often degrade the retinal structure associated with the patient.

\begin{figure}[t]
    \centering
    \includegraphics[width=0.95\columnwidth]{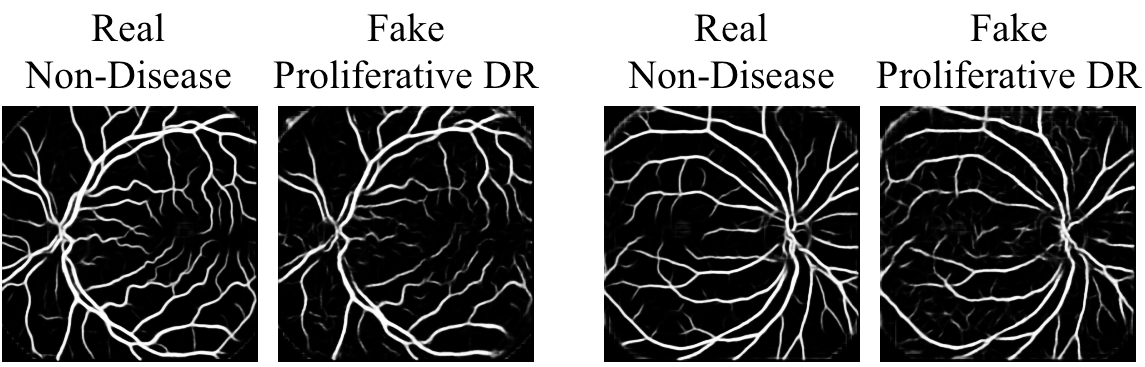}
    \caption{Vessel segmentation masks of input non-disease and output Proliferative} DR retinal images.
    \label{fig:segmentation_retinal}
    \vspace{-0.1in}
\end{figure}

\para{Comparing retinal vessel masks.}To further demonstrate identity preservation, we compare retinal vessel masks of the non-disease image and the generated image. The vascular patterns are known to aid in identification~\cite{costa_retinal_generation_2017}. We thus aim to investigate the \textit{similarity} between the vessel mask structure of the input non-disease image, and the generated image. If identity is preserved, we expect the vessel masks to be visually similar.
To extract the vessel mask, we train a U-Net based vessel segmentation model~\cite{unet_retinal} using DRIVE dataset~\cite{staal:2004-855}. Next, we apply the vessel extraction model to both the victim's original image, and the generated image. Paired examples are shown in Figure~\ref{fig:segmentation_retinal}. We can see the vessel structure is mostly preserved in the generated images. To obtain a quantitative measure, we compute MSSIM scores between the input and output vessel structures. For Proliferative DR, we obtain an MSSIM score of $0.71$. Score is not a perfect $1.0$, because blood vessels can swell and leak in the case of Proliferative DR. We also examined MSSIM scores for Severe DR, and obtain higher scores of 0.89, which is expected as this is a less advanced stage.

\begin{table}[t]
    \def\arraystretch{1.2}
    \caption{Disease injection ($R_d$) and identity preserving ($R_i$) performance of \agan{}.}
    \small
    \begin{center}
        \begin{tabular}{l|c|c}
            \hline
            \textbf{Dataset} & $\boldsymbol{R_d}$ &$\boldsymbol{R_i}$ \\
            \hline
            \textbf{Cardiomegaly}  & 82.8\% &95.2\%\\
            \textbf{Effusion}  & 95.7\% &94.4\%\\
            \textbf{Proliferative DR}  & 100\% & 88.4\%\\
            \hline
        \end{tabular}
    \end{center}
    \label{tab: jekyll_results}
    \vspace{-0.1in}
\end{table}

\begin{table}[t]
    \def\arraystretch{1.2}
    \caption{Disease injection ($R_d$) and identity preserving ($R_i$) performance of progressively injecting a disease.}
    \begin{center}
        \begin{tabular}{l|c|c}
            \hline
            \textbf{Dataset} & $\boldsymbol{R_d}$ &$\boldsymbol{R_i}$ \\
            \hline
            \textbf{Severe DR}  & 99.6\% & 97.5\%\\
            \textbf{Proliferative} DR & 100\% & 88.4\%\\
            \hline
        \end{tabular}
    \end{center}
    \label{tab: attackGAN__progressive_injection_results}
    \vspace{-0.2in}
\end{table}

\vspace{-1ex}
\subsubsection{Sustaining the Attack Over Time}\label{sec: sustaining_attack_over_time}\hfill\\
One of our attack goals is to enable repeated attacks using \agan{} over time. We consider a setting where attacker wants to mimic the natural progression of a disease condition, \eg condition worsening over time. As discussed in Section~\ref{sec:methodology}, we consider two scenarios.

\para{Disease stage data available.} We demonstrate the efficacy of this approach using the retinal dataset. In addition to Proliferative DR, we demonstrate injection of an earlier stage, namely Severe DR. Disease injection ($R_d$) and identity preservation ($R_i$) rates for this stage is presented in Table~\ref{tab: attackGAN__progressive_injection_results}. \agan{} is able to achieve over 99\% disease injection rate, and over 88\% identity preservation rate for both Proliferative and Severe stages. Sample images are provided in Appendix~\ref{appendix: supplementary_images} (Figure~\ref{fig:appendix_progressive_retinal}).

\begin{figure}[t]
    \centering
    \includegraphics[width=0.95\columnwidth]{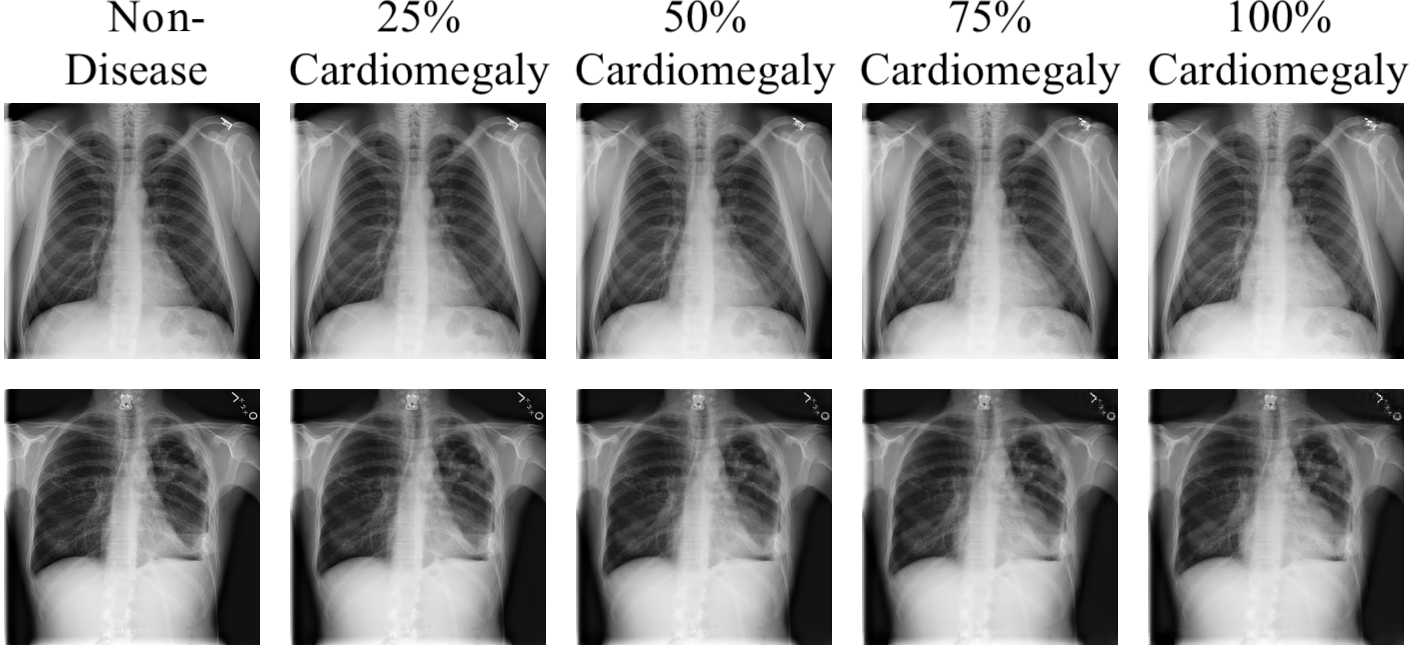}
    \caption{Examples of progressive injection for Cardiomegaly using linear interpolation.}
    \label{fig:progressive_injection__cardio_interpolation}
    \vspace{-0.2in}
\end{figure}

\para{Disease stage data unavailable.} In this case, attacker has no data to train \agan{} to target different stages. We use linear interpolation (Section~\ref{sec:methodology}) to produce intermediate stages of Cardiomegaly. Figure~\ref{fig:progressive_injection__cardio_interpolation} illustrates one example, where we produce 3 intermediate stages of Cardiomegaly using degree of disease injection, $\alpha \in [0.25,0.50,0.75]$. We see that this simulates enlargement of heart over time, while preserving patient identity. We further verified several such examples (using different victim images) with a doctor experienced in analyzing chest X-rays (who also participated in our user study in Section~\ref{ssec:userstudy}). The doctor confirmed that these intermediate stages are indeed plausible. To obtain a quantitative measure for progressive disease injection, we further apply the evaluation disease classifier, $C_{e}^{d}$ on all intermediate stages of all victims. The classifier produces a monotonically increasing probability of image being in the Cardiomegaly class, when we increase $\alpha$ (as shown in Figure~\ref{fig:progressive_cardio_linear_interpolation_line_curve} in Appendix~\ref{appendix: supplementary_images}). 

Overall, in both settings, \agan{} is able to mimic progression of a disease.

\begin{table*}[!htb]
    \caption{Disease Injection ($R_d$) and Identity Preservation ($R_i$) performance for ablation study and alternative architectures. }
    \vspace{-5mm}
    \def\arraystretch{1.2}
    \scriptsize
    \begin{center}
        \begin{tabular}{l|c|c|c|c|c|c|c|c|c|c|c|c|c|c|c}
            \hline
            \multirow{2}{*}{\textbf{Datasets}} & \multicolumn{2}{c|}{\multirow{2}{*}{\textbf{Jekyll}}} & \multicolumn{2}{c|}{\multirow{2}{*}{\textbf{Only} $\boldsymbol{L_{cycle}}$}} & \multicolumn{2}{c|}{$\boldsymbol{L_{disease}}$} & \multicolumn{2}{c|}{$\boldsymbol{L_{identity}}$}  & \multicolumn{2}{c|}{\multirow{2}{*}{\textbf{Pix2Pix}}} & \multicolumn{2}{c|}{\multirow{2}{*}{\textbf{IPCGAN}}} & \multicolumn{2}{c}{\multirow{2}{*}{\textbf{StarGAN}}}\\ 
            &\multicolumn{2}{c|}{} & \multicolumn{2}{c|}{} & \multicolumn{2}{c|}{\textbf{removed}} & \multicolumn{2}{c|}{\textbf{removed}} &\multicolumn{2}{c|}{}&\multicolumn{2}{c|}{}&\multicolumn{2}{c}{} \\\cline{2-15}
            \textbf{} & \multicolumn{1}{c|}{$R_d$} & \multicolumn{1}{c|}{$R_i$} & \multicolumn{1}{c|}{$R_d$} & \multicolumn{1}{c|}{$R_i$} & \multicolumn{1}{c|}{$R_d$} & \multicolumn{1}{c|}{$R_i$} & \multicolumn{1}{c|}{$R_d$} & \multicolumn{1}{c|}{$R_i$} & \multicolumn{1}{c|}{$R_d$} & \multicolumn{1}{c|}{$R_i$} & \multicolumn{1}{c|}{$R_d$} & \multicolumn{1}{c|}{$R_i$} & \multicolumn{1}{c|}{$R_d$} & \multicolumn{1}{c}{$R_i$}\\
            \hline
            \textbf{Cardiomegaly} & \multicolumn{1}{c|}{82.8\%} & \multicolumn{1}{c|}{95.2\%} & \multicolumn{1}{c|}{45.2\%} & \multicolumn{1}{c|}{94.2\%} & \multicolumn{1}{c|}{49.4\%} & \multicolumn{1}{c|}{94.6\%} & \multicolumn{1}{c|}{75.7\%} & \multicolumn{1}{c|}{86.7\%} & \multicolumn{1}{c|}{62.2\%} & \multicolumn{1}{c|}{0.3\%} & \multicolumn{1}{c|}{36.1\%} & \multicolumn{1}{c|}{99.3\%} & \multicolumn{1}{c|}{74.6\%} & \multicolumn{1}{c}{79.8\%}\\
            \textbf{Effusion} & \multicolumn{1}{c|}{95.7\%} & \multicolumn{1}{c|}{94.4\%} & \multicolumn{1}{c|}{44.2\%} & \multicolumn{1}{c|}{96.9\%} & \multicolumn{1}{c|}{55.2\%} & \multicolumn{1}{c|}{97.1\%} & \multicolumn{1}{c|}{97.5\%} & \multicolumn{1}{c|}{91.4\%} & \multicolumn{1}{c|}{97.6\%} & \multicolumn{1}{c|}{0.08\%} & \multicolumn{1}{c|}{64.3\%} & \multicolumn{1}{c|}{99.3\%} & \multicolumn{1}{c|}{70\%} & \multicolumn{1}{c}{90.8\%}\\
            \textbf{Proliferative DR} & \multicolumn{1}{c|}{100\%} & \multicolumn{1}{c|}{88.4\%} & \multicolumn{1}{c|}{96.7\%} & \multicolumn{1}{c|}{85\%} & \multicolumn{1}{c|}{89.3\%} & \multicolumn{1}{c|}{93.8\%} & \multicolumn{1}{c|}{99.8\%} & \multicolumn{1}{c|}{82.2\%} & \multicolumn{1}{c|}{100\%} & \multicolumn{1}{c|}{0.1\%} & \multicolumn{1}{c|}{2.4\%} & \multicolumn{1}{c|}{98.5\%} & \multicolumn{1}{c|}{87.9\%} & \multicolumn{1}{c}{93.1\%}\\
            \hline
        \end{tabular}
    \end{center}
    \label{tab: ablation_and_alternative_results}
    \vspace{-0.2in}
\end{table*}

\vspace{-1ex}
\subsubsection{Effectiveness of Jekyll when using alternative architectures}
\label{sec:alternative-arch}
We evaluate alternative neural network architectures to implement \agan{}. Two existing image-to-image translation models, StarGAN and IPCGAN are adapted to fit into \agan{}'s framework (discussed earlier in Section~\ref{sec:methodology}). In addition, as a baseline, we evaluate how a basic image-to-image translation framework would perform on our attack dataset. For this baseline, we choose the Pix2Pix~\cite{isola2017image} model, an image-to-image translation GAN that requires a paired dataset. Training objectives for Pix2Pix include an adversarial loss term, and an $L1$ loss term that minimizes the differences between the generated image and the targeted image in the pair. There are no other loss terms to enforce disease injection or identity. Since we do not have paired data, we use random pairs of non-disease and disease image to train Pix2Pix.

Table~\ref{tab: ablation_and_alternative_results} presents disease injection and identity preservation rates for all architectures---StarGAN, IPCGAN and Pix2Pix. We note that all architectures produce high quality images. However, there is variance in their attack effectiveness, and our original architecture outperforms all of them. \textit{StarGAN} performs reasonably well, achieving over $70\%$, and $79\%$, disease injection and identity preservation rate, respectively for all datasets. We suspect that using a single generator-discriminator pair in StarGAN (compared to using 2 pairs in our framework) is negatively impacting its translation performance. On the other hand, \textit{IPCGAN} achieves high identity preservation rate of over 99\% for all datasets, but disease injection rates are lower, with the retinal dataset showing only over 2\% performance. While adjusting the weights of the loss terms, we observe that it is necessary to upweight the identity loss term in order to obtain passable image quality. This likely explains the high $R_i$ rate. Any attempt to raise the weight of the disease loss term resulted in lowered image quality. Also note that IPCGAN does not include a cycle loss term. Lastly, \textit{Pix2Pix} exhibits high disease injection performance, but almost completely fails to preserve identity. This is expected given that Pix2Pix expects paired data (which is not available) and includes no train objectives to preserve identity (cycle loss or identity loss).

\vspace{-1ex}
\subsubsection{Ablation Study}
Table~\ref{tab: ablation_and_alternative_results} presents results of our ablation study. In all experiments, we retain the adversarial loss and cycle loss ($L_{cycle}$) terms, as they are basic building blocks of \agan{}'s training objectives. On removing the disease loss term ($L_{disease}$), disease injection rates decrease for all three datasets, with X-ray datasets exhibiting $R_d$ below $55\%$. Hence, the adversarial loss term on its own is insufficient for effective disease injection all the time, and we need a disease loss term. However, since disease-specific perturbations are no longer enforced, the patients' structural integrity is easily preserved, and the identity preservation rate ($R_i$) is high for all diseases.

To preserve identity, \agan{} relies on the cycle loss ($L_{cycle}$) and identity loss ($L_{identity}$) terms. When $L_{identity}$ is removed, $R_i$ decreases for all three diseases---an average of 5.8\% decrease for chest diseases, and 6.2\% decrease for Proliferative DR. However, the cycle loss term helps to avoid significant drop in identity preservation rate. We also observe noise artifacts, along with blurring in images generated without the identity loss. Therefore, the perceptual nature of identity loss contributes to image quality as well.

When both $L_{disease}$, and $L_{identity}$ are removed, \agan{} only uses the cycle loss term, and is equivalent to the vanilla CycleGAN translation model~\cite{zhu2017unpaired}. The disease injection rate drops significantly for both X-ray datasets (below $46\%$), while $R_i$ is high across all datasets. This is because when disease perturbations are not enforced, it is easy to achieve high $R_i$. This shows that a vanilla CycleGAN model is not sufficient for our attack.

To summarize, $L_{disease}$ is crucial to achieve high disease injection rate. But substantial disease perturbations can hurt the identity of the patient, and therefore $L_{identity}$ helps to balance things out, and preserve identity.

\vspace{-1ex}
\subsection{Evaluation by Medical Professionals}
\label{ssec:userstudy}
\vspace{-1ex}

\noindent To understand \agan{}'s ability to mislead medical professionals, we recruit medical professionals experienced in Chest X-ray diagnostics and conduct a user study. We investigate two key questions: (1) \textit{Can we convince medical professionals that \agan{} generated images contain the targeted disease condition?} (2) \textit{Can medical professionals distinguish between real images and \agan{} generated (fake) images?} We use the Chest X-ray dataset, and Cardiomegaly as the attacker targeted disease condition. Due to logistical difficulty in recruiting professionals with adequate experience in assessing retinal fundus imagery, we limit the study to the Chest X-ray dataset. Note that our study does not impact any real patients. Prior to conducting our study, we submitted a human subject protocol and received approval from our local IRB board.

We recruited three qualified medical practitioners, with extensive experience in evaluating chest X-Rays. Two evaluators are resident physicians in the Internal Medicine Department at a hospital. The third evaluator is a senior resident physician in the General Medicine Department. All three evaluators work at different hospitals. A pre-study questionnaire confirmed that all three practitioners have experience with analyzing chest X-rays.

The study requires both fake images with disease condition, and real images with and without disease condition. We sampled disease images (real and fake) from those having a disease probability higher than $0.8$ when evaluated by the evaluation disease classifier ($C_{e}^{d}$). Further, we observe that real X-rays contain watermarks (usually in upper right corner). \agan{} images contained slightly blurry reproduction of watermarks in the original victim's image. To ensure a fair evaluation, where evaluators would focus on the biomedical content, rather than quality of watermarks, we apply a simple automated post-processing step to sharpen the watermarks in fake images. \footnote{We use a similar approach from Section~\ref{sec:methodology} (on sustaining attack over time) and apply a linear interpolation scheme in a localized region containing the watermark, using the original non-disease image and the generated image.}

\vspace{-1ex}
\subsubsection{User Study Part 1: Evaluating Disease Injection}
In this task, given an X-ray image, evaluator has to determine whether it contains the disease condition of Cardiomegaly. Evaluators are offered three choices: (1) `Disease', image indicates the disease condition, (2) `No disease', no disease condition is observed, and (3) `Other', if evaluator is unable to choose (1) or (2) for some reason. We did not enforce a time limit to evaluate a given image. Figure~\ref{fig:user_study_t1} in Appendix~\ref{appendix: user_study} shows our survey page.

Each evaluator is shown a total of 150 images, out of which 100 are real, and the remaining 50 are fake images containing Cardiomegaly. Among the 100 real images, 50 contain no disease condition, and the other 50 contain Cardiomegaly. Evaluators are not aware of the presence of fake images. Each of the three evaluators independently evaluated all 150 images, in a total of 5 sessions (30 images per session), providing us a total of 450 judgements. Out of these 450 judgements, we discarded 5 where the evaluators were uncertain about their decision (`Other' class).\footnote{For these 5 uncertain cases, evaluators gave us feedback that some images were rotated, or had poor visibility in the heart region, making it harder to arrive at a decision.}

\begin{table}[t]
    \def\arraystretch{1.2}
    \caption{Accuracy of evaluators when (1) identifying real X-rays with symptoms of Cardiomegaly (Fleiss' $\kappa=0.62$ ($p<10^{-6}$)), and (2) distinguishing between real and fake X-rays.}
    \begin{center}
        \begin{tabular}{l|c|c}
            \hline
            \multirow{3}{*}{\textbf{Evaluator}} & \multicolumn{2}{c}{\textbf{Accuracy}} \\ \cline{2-3}
            & \multicolumn{1}{c|}{\textbf{Study Part 1}} & \multirow{2}{*}{\textbf{Study Part 2}} \\
            & \textbf{(Real X-rays only)} \\
            \hline
            \textbf{Evaluator} 1 & 91.9\%  & 48.0\% \\
            \textbf{Evaluator} 2 & 86.6\% & 48.0\% \\
            \textbf{Evaluator} 3 & 91.8\% & 55.0\% \\
            \hline           
            \textbf{Average}     & 86.8\% & 50.3\% \\
            \textbf{Majority}    & 90.9\% & 47.0\% \\
            \hline
        \end{tabular}
    \end{center}
    \label{tab:evaluator-accuracy-baseline}
    \vspace{-0.2in}
\end{table}

We start by investigating the baseline accuracy of the evaluators for this task, by comparing their judgments on the 100 real images.
Second column of Table~\ref{tab:evaluator-accuracy-baseline} shows the individual accuracy of evaluators,
as well as the average and majority accuracy.
The average accuracy is computed
by taking the mean of the individual accuracies,
while majority accuracy is computed
by taking the majority judgments of the three evaluators.
On average, accuracy of the evaluators is 0.868,
while when taking majority judgments, the accuracy rises to 0.909. Corresponding false-positive and false-negative rates are provided in Tables~\ref{tab: evaluator-fp} and~\ref{tab: evaluator-fn} (in Appendix).
The inter-annotator agreement, when computed using Fleiss' $\kappa$ metric,
is $\kappa=0.62$ ($p<10^{-6}$) indicating substantial agreement. %

\begin{table}[t]
    \def\arraystretch{1.2}
    \caption{Percentage of real X-ray images and fake X-ray images judged by evaluators to contain symptoms of Cardiomegaly.}
    \begin{center}
        \begin{tabular}{l|c|c}
            \hline
            \textbf{\# Evaluators voting `Disease'} & \textbf{Real Images} & \textbf{Fake Images} \\
            \hline
            \textbf{At least 1 evaluator} & 98\% & 98\% \\
            \textbf{At least 2 evaluators} & 88\% & 80\% \\
            \textbf{All 3 evaluators} & 60\% & 60\% \\
            \textbf{0 evaluators} & 2\% & 2\% \\
            \hline
        \end{tabular}
    \end{center}
\label{tab:evaluator-disease-judgements}
\vspace{-0.2in}
\end{table}

Next, we analyze the judgments of evaluators
on the 50 fake images (with Cardiomegaly) to understand the ability of \agan{}
in successfully injecting realistic looking disease conditions.
Column 3 of Table~\ref{tab:evaluator-disease-judgements}
shows the percentage of fake images that were judged to contain
diseased symptoms by zero, at least one, at least two or by all three evaluators.
We can see that for 60\% of the cases all evaluators agreed
that the fake images contained symptoms of the disease,
while in overall 80\% cases, the majority of the evaluators agreed
that the fake images contained symptoms of the disease.

Finally, we analyze if there is any statistical difference between judgements of real and fake disease images. Second column of Table~\ref{tab:evaluator-disease-judgements} also shows
the percentage of real images
that were judged to contain disease symptoms
by zero or more evaluators.
To understand whether this represented statistically significant differences,
we compared the ratio of the number of disease vs non-disease judged images
(according to the majority) using Pearson's Chi-squared test (a parametric test) and
Fisher's Exact Test (a non-parametric test).
When comparing the differences using Pearson's Chi-squared test
we obtain $\chi^2=0.669$ ($p=0.4132$).
Similarly, when using Fisher's Exact Test, we obtain
an odds ratio of $1.822$ ($p=0.4139$).
Since in both cases $p>>0.05$ we do not consider the differences to be
statistically significant. 

To summarize, medical professionals are convinced that \agan{} images contain the targeted disease (Cardiomegaly), and their judgements of real and fake disease images shows no statistical difference.

\vspace{-1ex}
\subsubsection{User Study Part 2: Detecting Fake Images}
In this task, evaluators are asked to mark a given image as `fake' or `real'. We described that `real' images are those produced by an X-ray imaging device, and `fake' are those generated by a computer algorithm. Evaluators were free to use any basis for their judgement. Again, we do not enforce any time limit to analyze the image. Figure~\ref{fig:user_study_t2} in Appendix~\ref{appendix: user_study} shows a screenshot of our survey page.

All three evaluators are given 50 real images with disease,
and 50 fake images (with disease) generated using \agan{}.
Third column of Table~\ref{tab:evaluator-accuracy-baseline} shows the individual accuracy of evaluators, as well as the average, and majority accuracy.
Overall, the accuracy when considering majority decision is low at 47\%.
The inter annotator agreement for the fake detection task is \mbox{$\kappa=0.024$} ($p = 0.674$),
when measured using Fleiss's $\kappa$ metric, indicating a lack of agreement among annotators.
Therefore, we conclude that evaluators are unable to accurately distinguish between real and fake images.

\begin{table*}[t]
    \def\arraystretch{1.2}
    \centering
    \caption{Performance of the Supervised MesoNet classifier and blind CSD SVM when detecting fake images generated by \agan{}. Precision and Recall are presented for the fake images generated by \agan{}.}
    \begin{center}
        \begin{tabular}{l|c|c|c|c|c|c|c|c|c}
        \hline

        \multirow{3}{*}{\textbf{Datasets}} & \multicolumn{6}{c|}{\textbf{Supervised: MesoNet}} & \multicolumn{3}{c}{\textbf{Blind: CSD-SVM}}\\ \cline {2-10}
        &\multicolumn{3}{c|}{\textbf{Jekyll Images}} &\multicolumn{3}{c|}{\textbf{Repurposed Jekyll Images}} &\multicolumn{3}{c}{\textbf{Jekyll Images}}  \\ \cline{2-4} \cline{5-7} \cline{8-10} 
        & \multicolumn{1}{c|}{\textbf{Accuracy}} & \multicolumn{1}{c|}{\textbf{Precision}} & \multicolumn{1}{c|}{\textbf{Recall}} & \multicolumn{1}{c|}{\textbf{Accuracy}} & \multicolumn{1}{c|}{\textbf{Precision}} & \multicolumn{1}{c|}{\textbf{Recall}} &
        \multicolumn{1}{c|}{\textbf{Accuracy}} & \multicolumn{1}{c|}{\textbf{Precision}} & \multicolumn{1}{c}{\textbf{Recall}} \\
        \hline
        \textbf{Cardiomegaly}  & 99.4\% &100\% &99\% &56.8\% &100\% &6\% & %
        N/A & N/A & N/A \\
        \textbf{Proliferative} DR  & 98.1\% &97\% &99\% &83.4\% &96\% &70\%&60.1\% &29\%&77\%\\
        \hline
        \end{tabular}
    \end{center}
    \label{tab:defense}
\end{table*}

\vspace{-1ex}
\section{Defending Against Attacks by \agan{}}
\label{sec:defenses}
\vspace{-1ex}
\noindent We investigate two methods to detect images generated by \agan{}: (1) \textit{blind detection}, and (2) \textit{supervised detection}

\para{Blind Detection.} In this scheme, the defender has no access to fake images and no knowledge of the attacker's generative model. However, a corpus of real images is available to the defender. These assumptions are realistic, but the problem setting is quite challenging. A viable defense approach in this scenario is to use anomaly detection to identify fake images.

There is limited existing work on blind detection of GAN generated images. We use a method proposed by Li et al.~\cite{li2018detection} that leverages disparities in color components between real and fake images (GAN generated images). The idea is that GAN training objectives typically place no explicit constrains to learn specific correlations among color components in the RGB space. This can result in inconsistencies when fake images are examined in other color spaces, namely HSV and YCbCr. This technique extracts color statistics of HSV and YCbCr color spaces of real images, and then uses these features to train a One-Class Support Vector Machine (SVM) for anomaly detection. The expectation is that the one-class SVM will flag fake images as anomalies in the (color) feature space. We refer to this method as \textit{CSD-SVM}.

Since this method requires color images, we only use the Proliferative DR dataset. X-ray images are in grayscale. We use an implementation of CSD-SVM provided by the authors.\footnote{\url{https://github.com/lihaod/GAN_image_detection}} Training dataset of CSD-SVM includes 232 real images with Proliferative DR. The testing dataset contains 700 real, and 703 fake images (both with Proliferative DR). The  SVM uses a radial basis kernel, where kernel coefficient ($\gamma$) is set to $0.1$, and the upper bound for the training error ($\nu$) is set to $0.12$. The hyper-parameters are estimated using 5-fold cross validation.

Table~\ref{tab:defense} shows detection accuracy, precision, and recall (for fake class). CSD-SVM is only able to achieve 60\% accuracy, at a precision and recall of 29\%, and 77\%, respectively. The low performance suggests that images generated by \agan{} are hard to detect using a blind scheme that analyzes color space disparity. %

\para{Supervised Detection.} Here we assume the defender has access to a corpus of both real and fake images. While this is a stronger defense assumption, it is important to investigate how well supervised schemes can detect fake images. Such schemes can be used along with a blind detection scheme to further improve detection performance. The basic idea is to build a supervised machine learning classifier to detect fake images. 

We use a high performing supervised classifier called MesoNet~\cite{afchar2018mesonet}, proposed recently by Afchar et al. The intuition behind \textit{MesoNet} is to examine image features at a \textit{mesoscopic} level, as micro (image noise) and macroscopic (semantic aspects) features may fail to distinguish between real and fake imagery. We use an implementation of MesoNet provided by the authors.\footnote{\url{https://github.com/DariusAf/MesoNet}}. The MesoNet model includes 4 inception blocks, where each block concatenates several convolutional layers of varying shapes together. Training and testing sets were constructed on disjoint sets of patients.

We apply MesoNet to the Cardiomegaly and Proliferative DR datasets. For the Cardiomegaly dataset, we train MesoNet on $4,920$ real (obtained via upsampling), and $4,359$ fake images, and test on a set of $443$ real and $375$ fake images. For the Proliferative DR dataset, we train MesoNet on 400 real and 403 fake images, and test on a balanced set of 300 real and 300 fake images.

Table~\ref{tab:defense} shows the detection performance of MesoNet. MesoNet performs well in detecting images generated by \agan{} with an accuracy of 99.4\% on the Cardiomegaly dataset, and 98.1\% accuracy on the Proliferative DR dataset. This is promising. However, note that the attacker can try to adapt the \agan{} model to bypass detection. Next, we present a countermeasure against this defense.

\para{Countermeasures Against Supervised Detection.} Our idea is to modify the attacker's GAN model to evade detection. If the attacker has white-box access to the defender model, one evasion strategy is to attach the defender's model to \agan{} as an additional discriminator. This will train the generator to produce fake images that can fool the defender's model. Instead, we assume a more realistic setting that does not require access to the defender model.

We tested different modifications to \agan{}, including retraining with different random seeds, retraining with different weights for the loss terms, changing the generator architecture by varying configuration of the deconvolution filters, and finally modifying the discriminator framework. Among all these modifications, we observe that changes to the discriminator framework results in successful evasion, which we describe in more detail below.

In a GAN, the generator primarily relies on the discriminator's feedback to generate fake images. \agan{} uses a PatchGAN discriminator that determines whether individual small patches of an image look real or fake. We modify \agan{} to include an additional discriminator that analyzes the image as a whole (instead of individual patches) to classify as fake or real. This additional discriminator is implemented using a single layer feed-forward network that takes the output of PatchGAN and produces a scalar score. The modified adversarial GAN objective is then updated to include both the direct PatchGAN output, as well as the output of the feed-forward network.
Such an additional discriminator serves to approximate the role of a defender model, which \agan{}'s generator can learn to fool while training. We refer to this modified version of \agan{} as \textit{Repurposed} \agan{}.

We then apply MesoNet trained on \agan{} images to images generated by \textit{Repurposed} \agan{}. Table~\ref{tab:defense} shows the detection performance. There is a significant drop in detection accuracy from 99.4\% to 55.6\% for the Cardiomegaly dataset, and from 98.1\% to 83.4\% for the Proliferative DR dataset. These results highlight the susceptibility of supervised classifiers to advanced attacks.

\vspace{-1ex}
\section{Conclusion}
\vspace{-1ex}
\noindent In this work, we investigate deep learning powered attacks against medical image diagnostics. We present \agan{}, a GAN-based framework that can translate a biomedical image of a patient to a new one that indicates an attacker chosen disease condition, while preserving the identity of the patient. Such translation attacks can lead to misdiagnosis by both medical professionals and machine learning algorithms. Additionally, \agan{} provides methods to enable repeated attacks against a victim by controlling the severity of the injected disease. We extensively evaluate attack success of \agan{} using both machine learning tools and a user study involving medical professionals. 
Lastly, we investigate defensive measures that aim to detect \agan{} generated images. We find that supervised detection approaches are promising but vulnerable to evasion tactics by advanced attackers. We hope our work encourages the community to pursue robust defensive measures.

\bibliographystyle{IEEEtran}
\bibliography{main}
\raggedbottom

\appendices

\section{Supplementary Images and Figures}
\label{appendix: supplementary_images}
\begin{figure}[H]
    \centering
    \includegraphics[width=0.98\columnwidth]{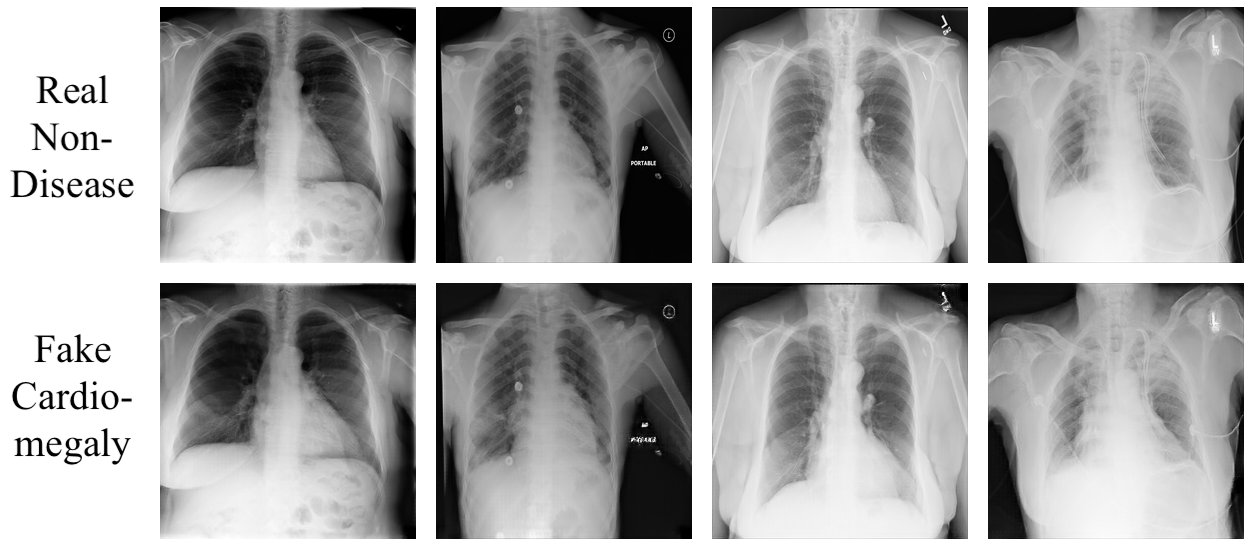}
    \vspace{0.1in}
    \includegraphics[width=0.98\columnwidth]{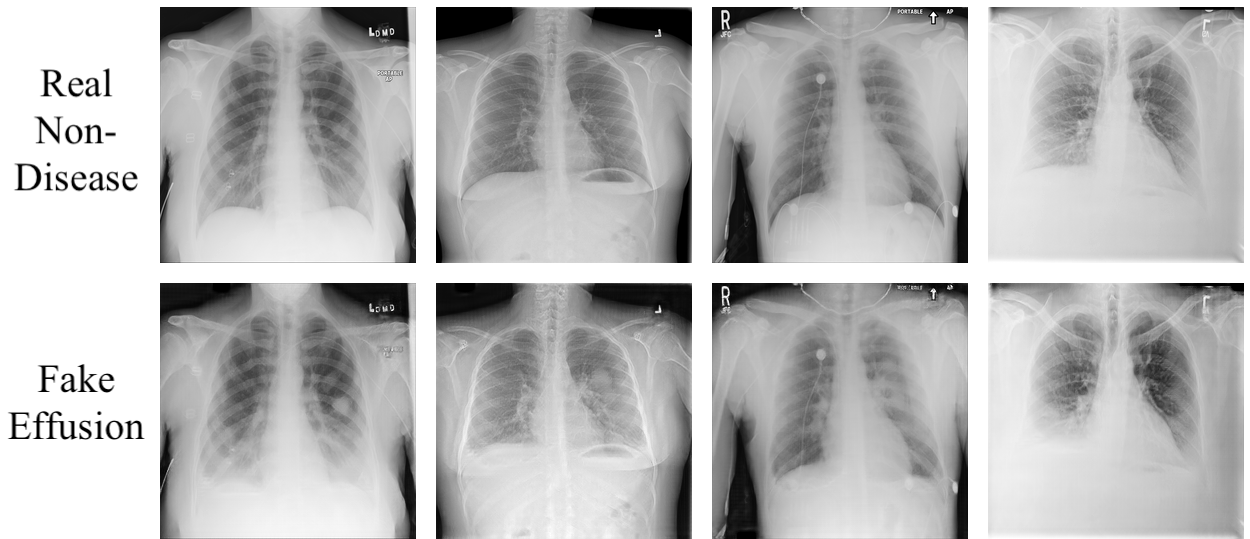}
    \vspace{0.1in}
    \includegraphics[width=0.98\columnwidth]{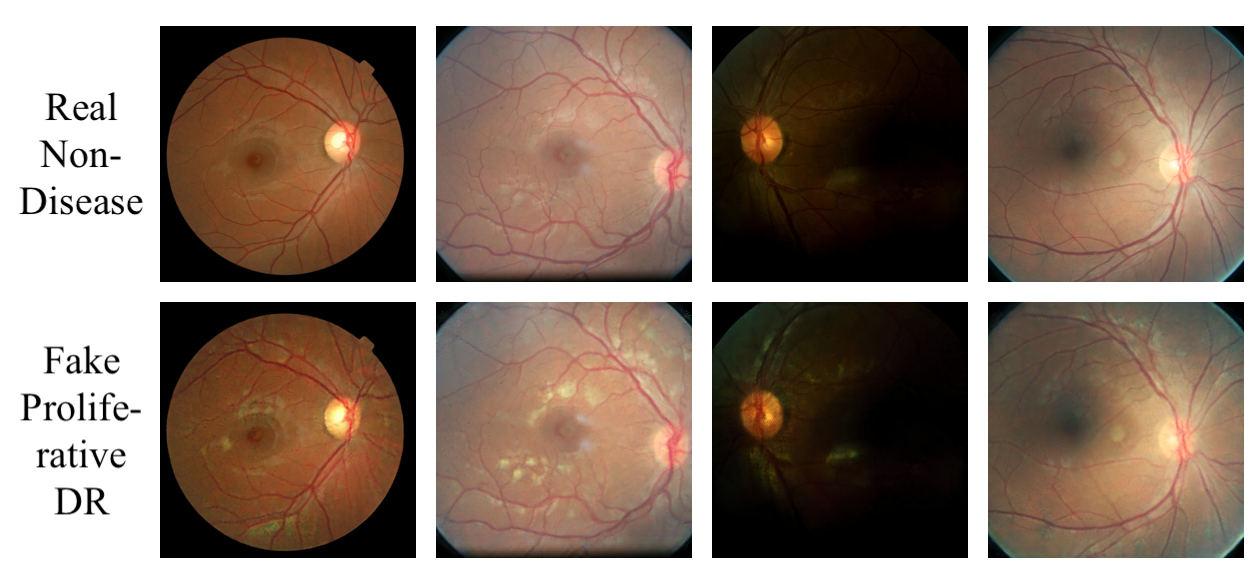}
    \caption{Disease Injection Samples for Cardiomegaly, Effusion and Proliferative DR}
    \label{fig:appendix_injection}
\end{figure}

\begin{figure}[H]
    \centering
    \includegraphics[width=0.98\columnwidth]{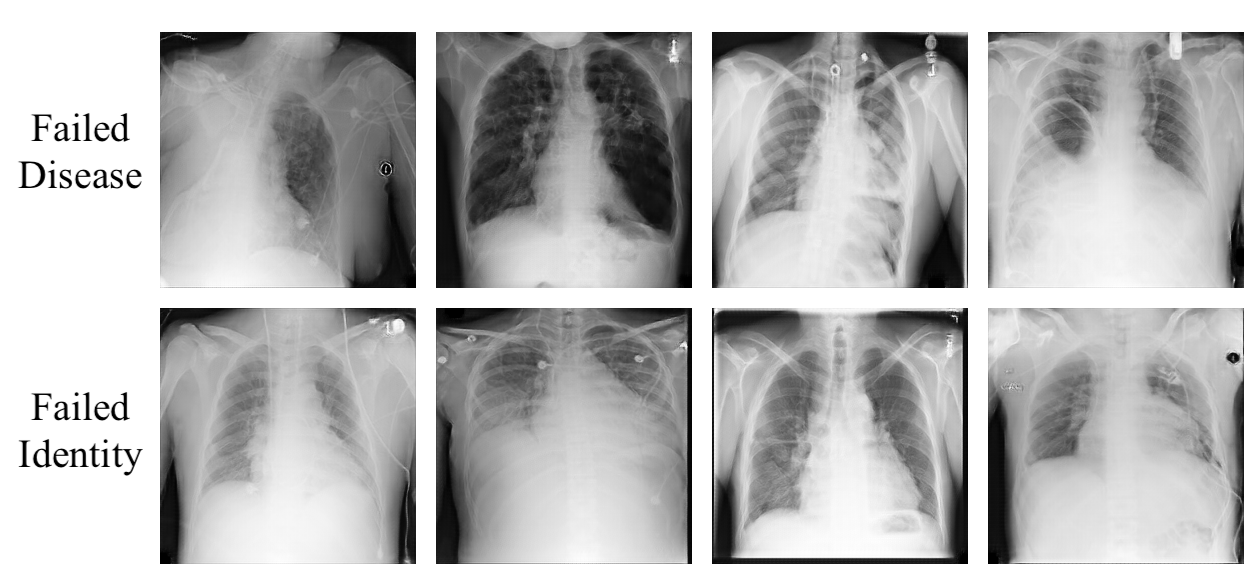}
    \caption{Failed Cardiomegaly Samples.}
    \label{fig:appendix_fail_cardio}
\end{figure}

\begin{figure}[H]
    \centering
    \includegraphics[width=0.98\columnwidth]{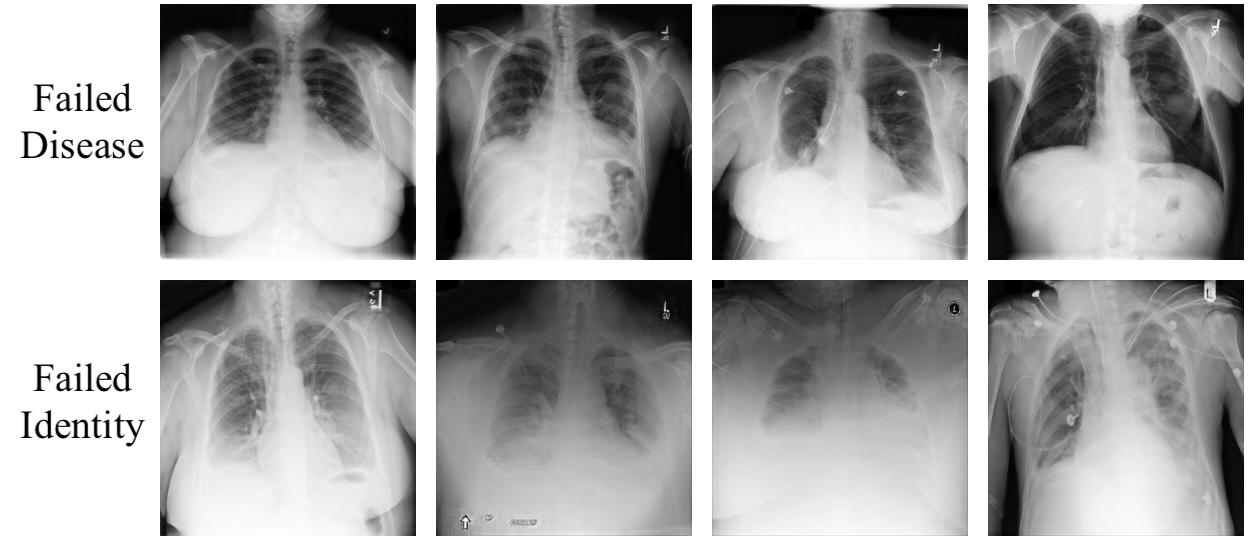}
    \caption{Failed Effusion Samples.}
    \label{fig:appendix_fail_effusion}
\end{figure}

\begin{figure}[H]
    \centering
    \includegraphics[width=0.98\columnwidth]{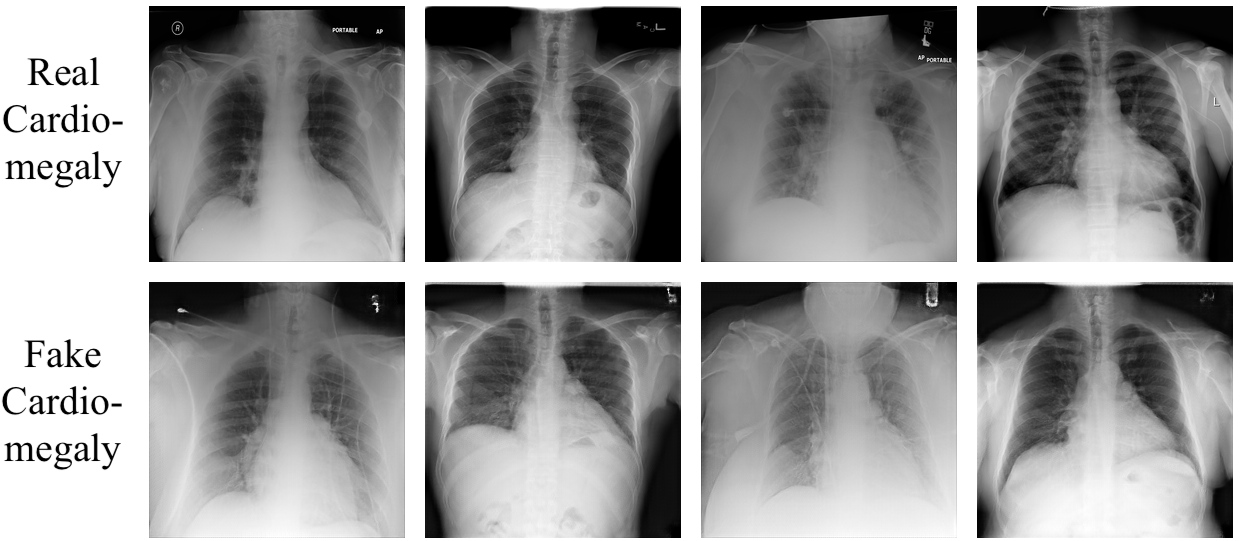}
    \vspace{0.1in}
    \includegraphics[width=0.98\columnwidth]{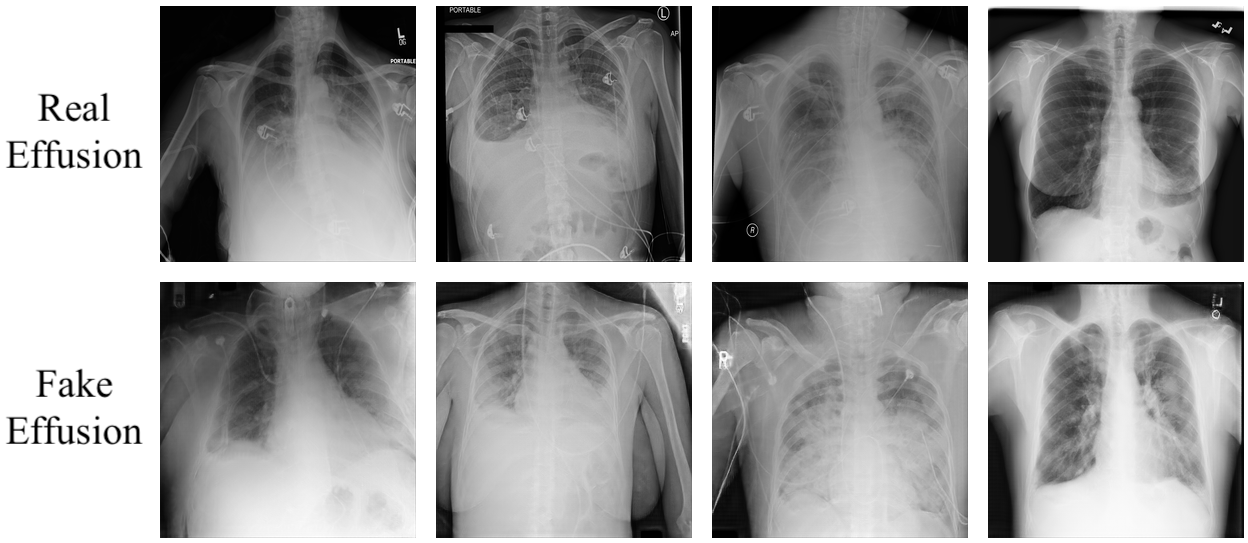}
    \vspace{0.1in}
    \includegraphics[width=0.98\columnwidth]{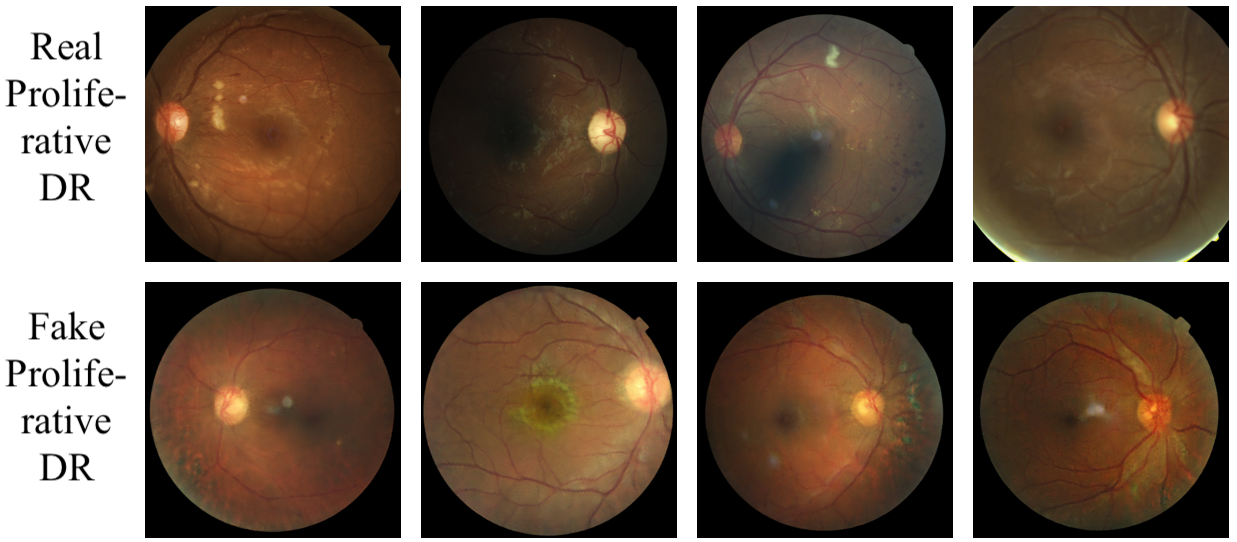}
    \caption{Real and Fake Samples for Cardiomegaly, Effusion and Proliferative DR}
    \label{fig:appendix_real_fake}
    \vspace{-0.2in}
\end{figure}

\begin{figure}[H]
    \centering
    \includegraphics[width=0.65\columnwidth]{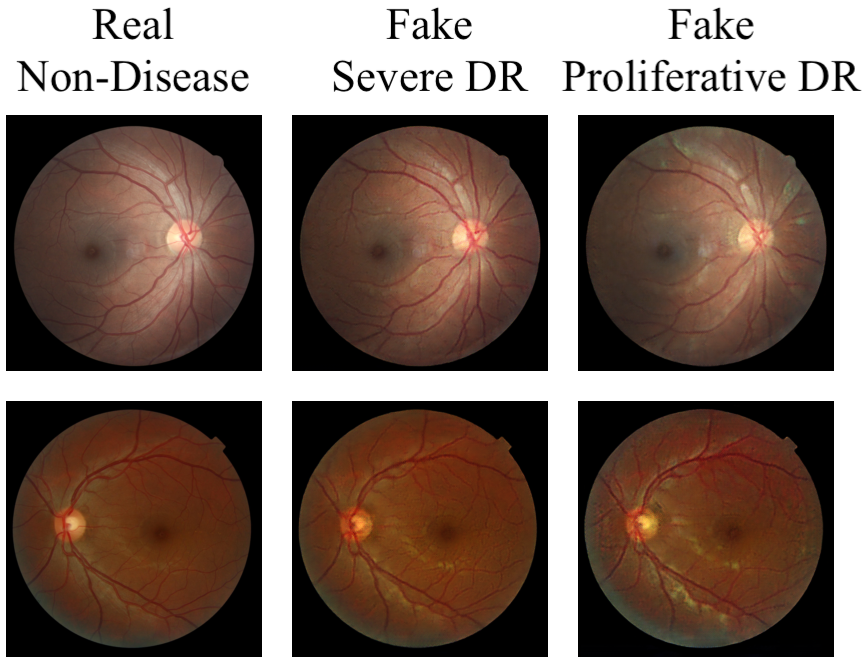}
    \caption{Examples of progressive injection for DR.}
    \label{fig:appendix_progressive_retinal}
\end{figure}

\begin{figure}[h]
    \centering
    \includegraphics[width=0.50\columnwidth]{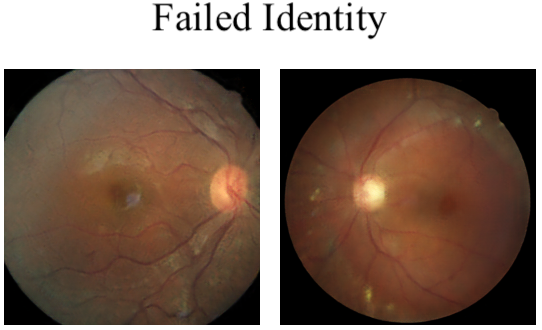}
    \caption{Examples of failure cases for Proliferative DR.}
    \label{fig:failed_proliferative_retinal}
\end{figure}

\begin{figure}[t]
    \centering
    \includegraphics[width=8cm]{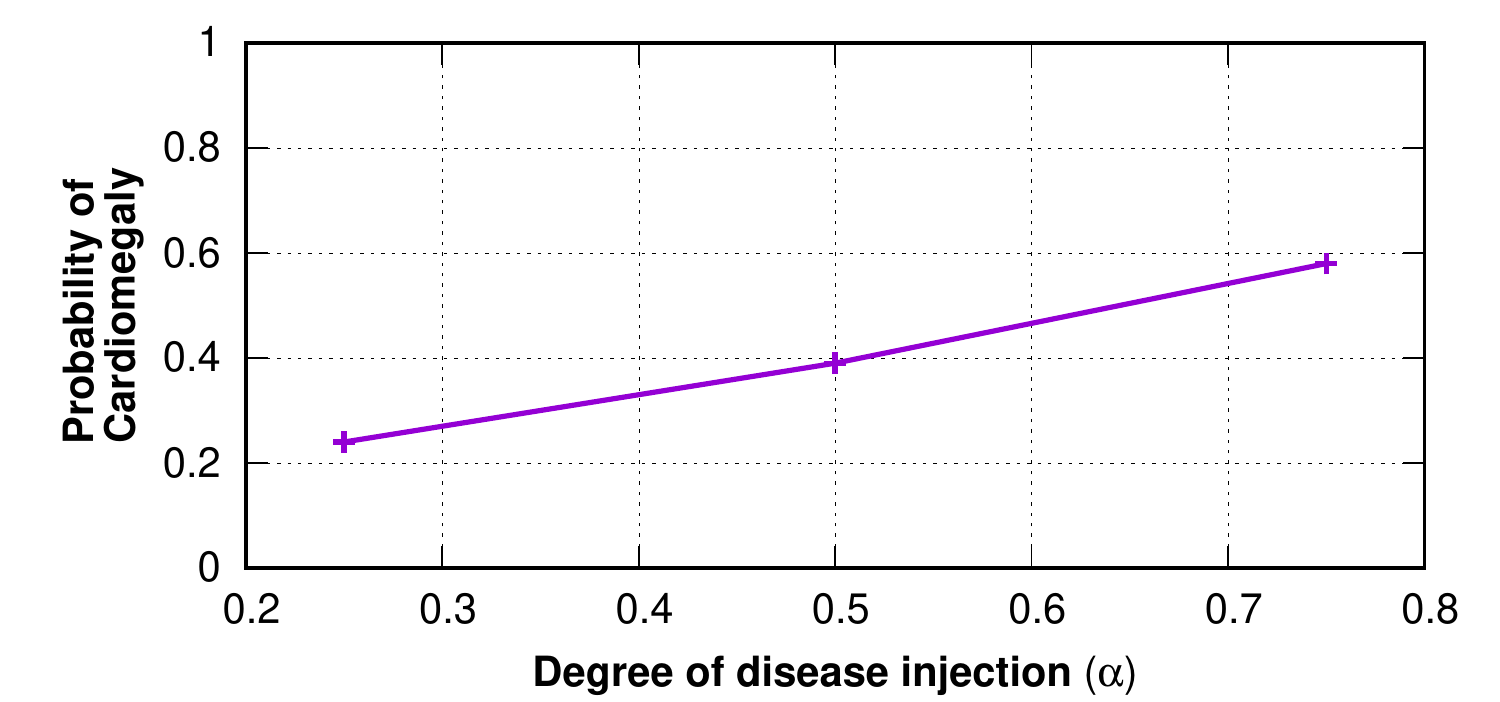}
    \caption{Monotonic increase in probability of Cardiomegaly as assigned by evaluation classifier $C_e^d$ when increasing degree of disease injection.}
    \label{fig:progressive_cardio_linear_interpolation_line_curve}
    \vspace{-0.1in}
\end{figure}

\section{User Study}\label{appendix: user_study}

\begin{figure}[H]
    \centering
    \includegraphics[width=0.92\columnwidth]{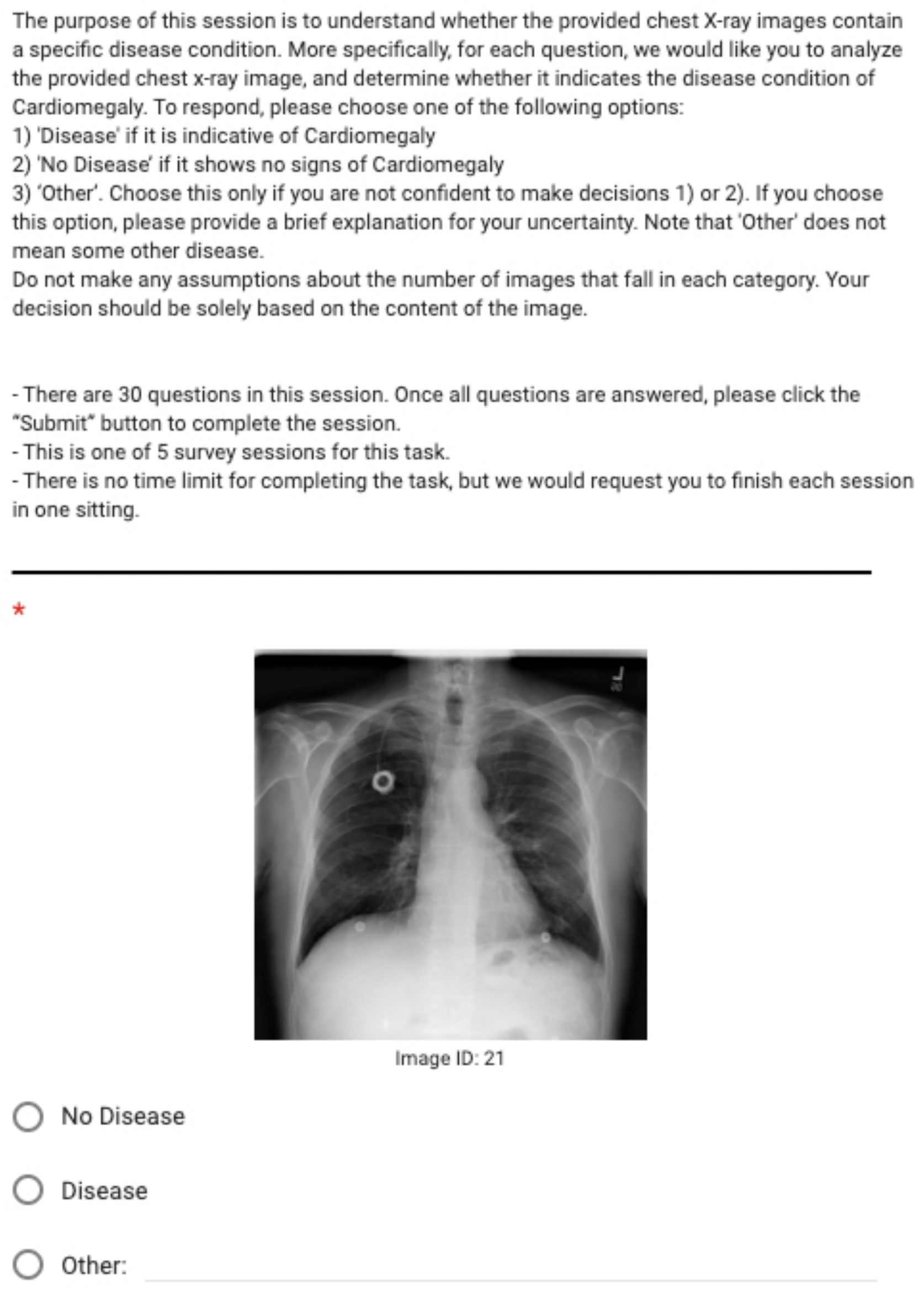}
    \caption{Screenshot of User Study Interface for Part 1.}
    \label{fig:user_study_t1}
\end{figure}

\begin{table}[H]
    \def\arraystretch{1.2}
    \caption{False positive rate of evaluators when (1) identifying real X-rays with symptoms of Cardiomegaly (Fleiss' $\kappa=0.62$ ($p<10^{-6}$)), and (2) distinguishing between real and fake X-rays.}
    \begin{center}
        \begin{tabular}{l|c|c}
            \hline
            \multirow{3}{*}{\textbf{Evaluator}} & \multicolumn{2}{c}{\textbf{False Positive Rate}} \\ \cline{2-3}
            & \multicolumn{1}{c|}{\textbf{Study Part 1}} & \multirow{2}{*}{\textbf{Study Part 2}} \\
            & \textbf{(Real X-rays only)} \\
            \hline
            \textbf{Evaluator} 1 & 2.0\%  & 9.0\% \\
            \textbf{Evaluator} 2 & 6.0\% & 7.0\% \\
            \textbf{Evaluator} 3 & 4.0\% & 8.0\% \\
            \hline           
            \textbf{Average}     & 4.0\% & 8.0\% \\
            \textbf{Majority}    & 3.0\% & 5.0\% \\
            \hline
        \end{tabular}
    \end{center}
    \label{tab: evaluator-fp}
\end{table}

\begin{table}[H]
    \def\arraystretch{1.2}
    \caption{False negative rate of evaluators when (1) identifying real X-rays with symptoms of Cardiomegaly (Fleiss' $\kappa=0.62$ ($p<10^{-6}$)), and (2) distinguishing between real and fake X-rays.}
    \begin{center}
        \begin{tabular}{l|c|c}
            \hline
            \multirow{3}{*}{\textbf{Evaluator}} & \multicolumn{2}{c}{\textbf{False Negative Rate}} \\ \cline{2-3}
            & \multicolumn{1}{c|}{\textbf{Study Part 1}} & \multirow{2}{*}{\textbf{Study Part 2}} \\
            & \textbf{(Real X-rays only)} \\
            \hline
            \textbf{Evaluator} 1 & 6.0\%  & 43.0\% \\
            \textbf{Evaluator} 2 & 7.0\% & 45.0\% \\
            \textbf{Evaluator} 3 & 14.2\% & 37.0\% \\
            \hline           
            \textbf{Average}     & 9.1\% & 41.7\% \\
            \textbf{Majority}    & 6.0\% & 48.0\% \\
            \hline
        \end{tabular}
    \end{center}
    \label{tab: evaluator-fn}
\end{table}

\begin{figure}[H]
    \centering
    \includegraphics[width=0.92\columnwidth]{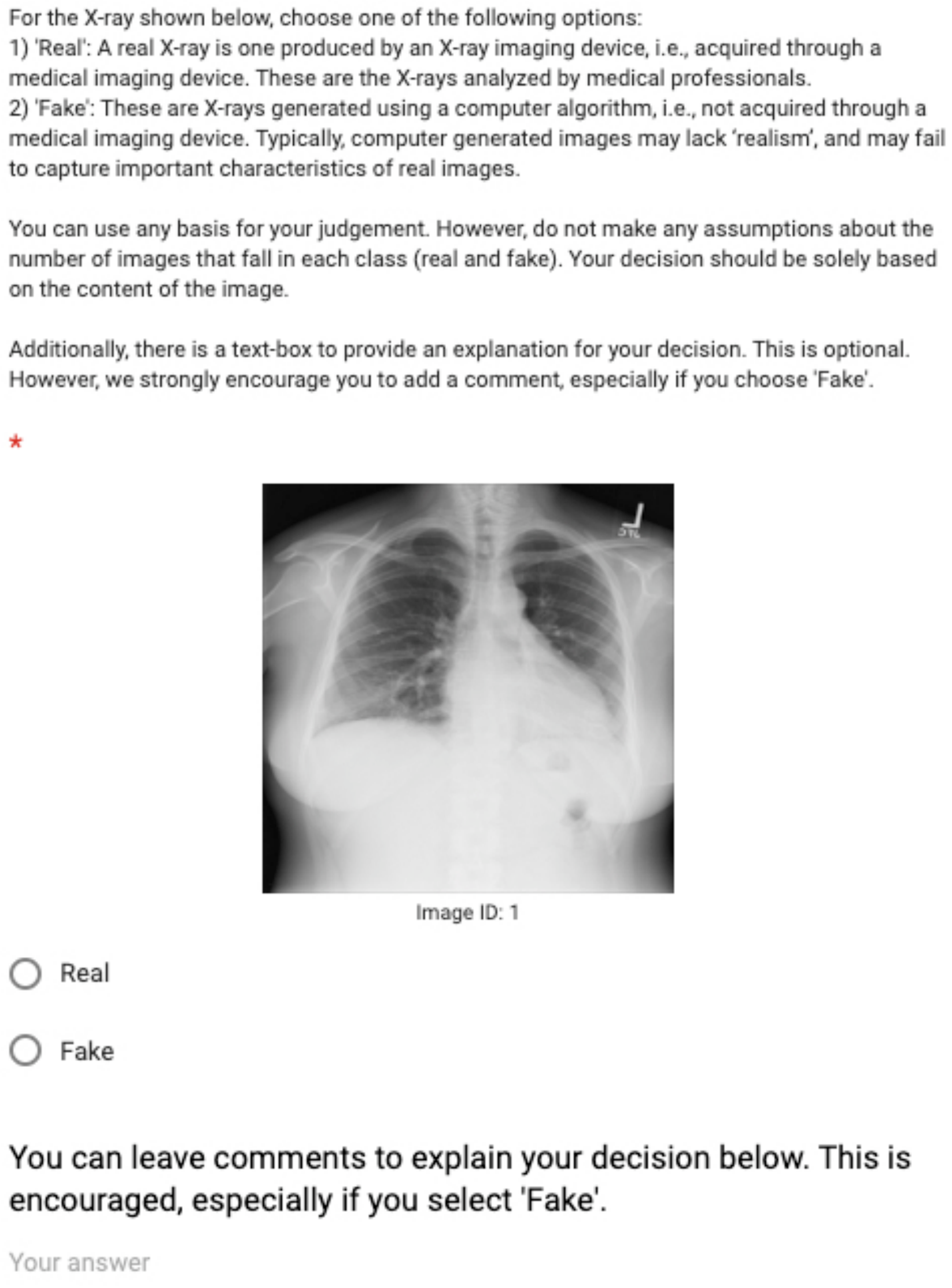}
    \caption{Screenshot of User Study Interface for Part 2.}
    \label{fig:user_study_t2}
\end{figure}

\section{Models and Datasets}
\label{appendix: models_and_datasets}
\subsection{Jekyll}
\begin{table*}
    \small
    \centering
    \caption{Generator and Discriminator architectures.}
    \begin{tabular}{lcccccc}
        \hline
        \multicolumn{7}{c}{\textbf{Generator Architecture}} \\ \hline
        Layer Type & \# of channels & Filter Size & Stride & Normalization & Activation & Output Shape \\ \hline
        padding & 3 & - &- &-  &-  & (262,262,3) \\ \hline
        conv2d & 64 & $7\times 7$ & 1 & instance\_norm & relu & (256,256,64) \\ \hline
        conv2d & 128 & $3\times 3$ & 2 & instance\_norm & relu & (128,128,128) \\ \hline
        conv2d & 256 & $3\times 3$ & 2 & instance\_norm & relu & (64,64,256) \\ \hline
        residual block & 256 &-  &-  &-  &-  & (64,64,256) \\ \hline
        residual block & 256 &-  &-  &-  &-  & (64,64,256) \\ \hline
        residual block & 256 &-  &-  &-  &-  & (64,64,256) \\ \hline
        residual block & 256 &-  &-  &-  &-  & (64,64,256) \\ \hline
        residual block & 256 &-  &-  &-  &-  & (64,64,256) \\ \hline
        residual block & 256 &-  &-  &-  &-  & (64,64,256) \\ \hline
        residual block & 256 &-  &-  &-  &-  & (64,64,256) \\ \hline
        residual block & 256 &-  &-  &-  &-  & (64,64,256) \\ \hline
        residual block & 256 &-  &-  &-  &-  & (64,64,256) \\ \hline
        deconv2d & 128 & $3\times 3$ & 2 & instance\_norm & relu & (128,128,128) \\ \hline
        deconv2d & 64 & $3\times 3$ & 2 & instance\_norm & relu & (256,256,64) \\ \hline
        padding & 3 &-  &-  &-  &-  & (262,262,3) \\ \hline
        conv2d & 3 & $7\times 7$ & 1 & instance\_norm & tanh & (256,256,3) ) \\ \hline
        \multicolumn{7}{c}{\textbf{Discriminator Architecture}} \\ \hline
        conv2d & 64 & $4\times 4$ & 2 & instance\_norm & relu & (128,128,64) \\ \hline
        conv2d & 128 & $4\times 4$ & 2 & instance\_norm & relu & (64,64,128) \\ \hline
        conv2d & 256 & $4\times 4$ & 2 & instance\_norm & relu & (32,32,256) \\ \hline
        conv2d & 512 & $4\times 4$ & 1 & instance\_norm & relu & (32,32,512) \\ \hline
        conv2d & 1 & $4\times 4$ & 1 & - & - & (32,32,1) ) \\ \hline
    \end{tabular}
    \label{tab: generator architecture}
\end{table*}

\begin{table*}
    \centering
    \caption{Training configurations of \agan{} for different diseases.}
    \begin{tabular}{l|c|c|c|c|c|c|c}
        \hline
         \textbf{Dataset} & \textbf{$\boldsymbol{\lambda_{adv}}$} & \textbf{$\boldsymbol{\lambda_{disease}}$} & \textbf{$\boldsymbol{\lambda_{identity}}$} & \textbf{$\boldsymbol{\lambda_{cycle}}$} & \textbf{Epochs} & \begin{tabular}[c]{@{}c@{}}\textbf{Batch} \\ \textbf{size}\end{tabular} & \textbf{Optimizer} \\ \hline
        \textbf{Cardiomegaly} & 20 & 50 & 25 & 200 & 100 & 1 & Adam \\
        \textbf{Severe DR} & 5 & 5 & 20 & 200 & 150 & 1 & Adam \\
        \textbf{Proliferative DR} & 5 & 10 & 20 & 200 & 160 & 1 & Adam \\ \hline
    \end{tabular}
    \label{tab: attackgan training setting}
\end{table*}

\begin{table*}

    \def\arraystretch{1.2}
    \centering
    \caption{Dataset statistics (number of images) and training configurations for attack ($C_a^d$) and evaluation ($C_e^d$) disease classifiers.}
        \begin{tabular}{l|c|c|c|c|c|c|c|c|c}
            \hline
            \multirow{2}{*}{\textbf{Dataset}} & \multicolumn{3}{c|}{$\boldsymbol{C_a^d}$} & \multicolumn{3}{c|}{$\boldsymbol{C_e^d}$} & \multirow{2}{*}{\textbf{ Epochs}} & \multirow{2}{*}{\textbf{Batch Size}} & \multirow{2}{*}{\textbf{Optimizer}}\\ 
            \cline{2-7}
            & \multicolumn{1}{c|}{\textbf{Train}} & \multicolumn{1}{c|}{\textbf{Validation}} & \multicolumn{1}{c|}{\textbf{Test}} & \multicolumn{1}{c|}{\textbf{Train}} &\multicolumn{1}{c|}{\textbf{Validation}} & \multicolumn{1}{c|}{\textbf{Test}} & \multicolumn{1}{c|}{} & \multicolumn{1}{c|}{} \\ 
            \hline
            \textbf{Cardiomegaly}  &1,600 & 100  &999 &1,600& 100 &1,177 & 50 & 5 & Adam\\
            \textbf{Effusion}  &7,000 & 1,454  &1,500 &11,676& 2,502 &2,502 & 50 & 5 & Adam\\
            \textbf{Severe DR}   &1,419 & 100 &609 &1,462& 100 &627 & 20 & 32 & Adam\\
            \textbf{Proliferative DR}  &1,374 & 100 &590 &1,339& 100 &575 & 20 & 32 & Adam\\
            \hline
        \end{tabular}
    \label{tab: disease_classifiers_data}
\end{table*}

\begin{table*}
    \def\arraystretch{1.2}
    \centering
    \caption{Dataset statistics (number of images) and training configurations for attack ($C_a^i$) and evaluation ($C_e^i$) identity classifiers. Training configuration for both $C_a^i$ and $C_e^i$ identity classifiers are the same for any given dataset.}
    \begin{tabular}{l|c|c|c|c|c|c|c|c|c}
       \hline
        \multirow{2}{*}{\textbf{Dataset}} & \multicolumn{3}{c|}{$\boldsymbol{C_a^i}$} & \multicolumn{3}{c|}{$\boldsymbol{C_e^i}$} & \multirow{2}{*}{\textbf{ Epochs}} & \multirow{2}{*}{\textbf{Batch Size}} & \multirow{2}{*}{\textbf{Optimizer}}\\ 
        \cline{2-7}
        & \multicolumn{1}{c|}{\textbf{Train}} & \multicolumn{1}{c|}{\textbf{Validation}} & \multicolumn{1}{c|}{\textbf{Test}} & \multicolumn{1}{c|}{\textbf{Train}} &\multicolumn{1}{c|}{\textbf{Validation}} & \multicolumn{1}{c|}{\textbf{Test}} & \multicolumn{1}{c|}{} & \multicolumn{1}{c|}{} \\ 
        \hline
        \textbf{X-Ray}  &6,249 & 1,598 &1,945 &13,162& 3,369 &4,108 & 1000 & 16 & Adam\\
        \textbf{Retinal Fundus}  &3,928& 1,000 &3,072 &4,157& 1,000 &3,249 & 80 & 3 & Adam\\
        \hline
    \end{tabular}
    \label{tab: identity_classifiers_data}
\end{table*}

\para{Dataset preparation for Jekyll.}

\noindent\textit{NIH Chest X-ray dataset.} In order to effectively partition the dataset for both the attack and evaluation partitions, we decided upon a division that would allow construction of a reliable identity and disease classifier for the evaluation partition. To achieve this goal, we construct partitions on a patient level, rather than an image level. More specifically, we first pick 2,000 patients such that each of these patients holds 10 or more images. This ensures that we can build a reliable evaluation identity classifier. However, we also require an abundance of images for each disease, in order to build effective evaluation disease classifiers. We thus add another 4,805 patients to the evaluation partition, to allow for a large amount of images for each disease. This brings the evaluation partition to a total of 6,805 patients. All remaining 24,000 patients in the NIH Chest X-ray dataset are used to train \agan{}.\newline

\noindent\textit{Retinal Fundus images.} The retinal fundus image dataset only provides 2 images per patient; one image for the left eye, and one image for the right eye. This limitation simplifies that division of patients for the attack and evaluation partitions. More specifically, we divide the patients equally between the attack and evaluation partitions, with each receiving
44,351 patients.\newline

\para{Implementation details for alternative architectures.} 
\noindent\textit{StarGAN.} To fit StarGAN into the \agan{} framework, as discussed in section \ref{sec:alternative-arch}, we must use a disease classification loss, and a perceptual identity loss. The disease classification loss objective is determined using predictions from the auxiliary domain classifier. This domain classifier shares weights with the adversarial discriminator. This objective is optimized via a modified version of the binary-cross entropy loss function, called the focal loss \cite{lin2017focal}. Focal loss modifies cross-entropy to downweight the loss coming from easily classified samples, and upweight that coming from difficult samples. We use hyperparameter values of $\alpha = 1$ and $\gamma = 2$. Here, $alpha$ is simply a balancing weight, and $\gamma$ is the \textit{focusing parameter}, which determines the rate at which easily classified samples are downweighted in the loss \cite{lin2017focal}. The perceptual identity loss objective is determined using an external attack identity classifier. The architecture of this external identity classifier (built separately in PyTorch for StarGAN) is identical to that of the default \agan{} attack identity classifier. More specifically, the identity loss is computed using output extracted from the third pooling layer (precedes the fourth dense block in the DenseNet-121 architecture). The model is trained for 200,000 iterations with a batch size of 1 and the Adam optimizer. Learning rate is set to $10^{-4}$. We weight the loss objectives with weights $\lambda_{adversarial} = 20, \lambda_{disease}=5, \lambda_{identity}=0.1, \lambda_{gradient\_penalty}=10, \lambda_{cycle}=200$.\newline

\noindent\textit{IPCGAN.} To fit IPCGAN into the \agan{} framework, we must again introduce a disease classification loss, and a perceptual identity loss. IPCGAN already optimizes an age-loss, and we substitute the AlexNet age-classifier for an external disease classifier. Similarly, IPCGAN already optimizes a perceptual identity loss, and we substitute the identity classifier for an external identity classifier. Again, the identity loss is computed using output extracted from the third pooling layer (precedes the fourth dense block in the DenseNet-121 architecture). Finally, the generator input layer is changed to accept 256x256x3 input, instead of 128x128x3, as used in the original IPCGAN implementation. The model is trained for 200,000 iterations, with a batch size of 8 and the Adam optimizer. Learning rate is set to $10^{-4}$. We weight the loss objectives with weights $\lambda_{adversarial}=1,\lambda_{disease}=1,\lambda_{identity}=0.0005$.

\end{document}